\documentclass[10pt,preprint]{aastex}
\slugcomment{Prepared for publication in PASP}
\shorttitle{Faint Photometric Sequences}
\shortauthors{Saha et al.}
\begin{document}
\title{Faint BVRI Photometric Sequences in Selected Fields}
\author{A.~Saha\altaffilmark{1,2}, A.~E.\ Dolphin\altaffilmark{1,3}
and F.~Thim\altaffilmark{1,4}} 
\affil{NOAO, P.O.\ Box 26732, Tucson, AZ 85726}
\email{saha@noao.edu, adolphin@as.arizona.edu, thim@noao.edu}
\and
\author{B.~Whitmore}
\affil{Space Telescope Science Institute, 3700 San Martin Drive, Baltimore, 
MD 21218}
\email{whitmore@stsci.edu}
\altaffiltext{1}{NOAO is operated by the Association of Universities for 
Research in Astronomy, Inc.\ (AURA) under cooperative agreement with the 
National Science Foundation}
\altaffiltext{2} {Sabbatical Visitor, Indian Institute of Astrophysics, Koramangala, Bangalore 560034, India}
\altaffiltext{3}{present address: Steward Observatory, University of Arizona, 933 North Cherry Avenue, Tucson, AZ 85721}
\altaffiltext{4}{present address: Brandenburg GmbH, Technologiepark 19, 33100 Paderborn, Germany; thim@brandenburg-gmbh.de}
\begin{abstract}
The results from work done to extend the Johnson-Cousins \textit{BVRI} 
photometric standard sequence to faint levels of $V \sim 21$ mag in compact 
fields is presented. Such calibration and extension of sequences is necessary 
to fill a calibration gap, if reliable photometry from modest aperture 
telescopes in space (e.g.\ \textit{HST}), or terrestrial telescopes with apertures 
exceeding 4-m is to done. Sequences like the ones presented here, which cover 
a large range in brightness as well as color, will allow photometric 
calibration to be done efficiently, as well as for such work to be less prone 
to systematic sources of error. 

Photometry of stars in approximately $10 \times 10$ arc-minute fields around 
3 globular clusters, NGC~2419, Pal~4 and Pal~14 are presented from data 
acquired over several photometric nights. In each field, 
several stars are measured in $B, V, R$ and $I$ passbands, with standard 
errors in the mean less than 0.015 mag from random errors, to levels fainter 
than $V = 21$ mag. It is shown that standard errors in the mean from 
systematic errors when tying to the Landolt standards on the Johnson-Cousins 
system are typically well below 0.01 mag in all 4 bands 
(except for $B$ in NGC~2419, and $R$ in Pal~4), thus justifying the claim 
that these fields have been correctly calibrated.

The primary context for the work presented here is that parts of these 
fields were observed repeatedly by the Wide Field Planetary Camera-2 (WFPC2) 
of the \textit{HST}, and thus these newly calibrated sequences can be used to 
retro-actively calibrate WFPC2 at over various times of its operating 
life. In the past, WFPC2 data have had typical photometric zero-point 
uncertainties of a few hundredths of a magnitude, largely due to a lack of 
suitable standard stars. The sequences presented 
here have standard errors at the 0.01 mag level. They agree at the 0.02 mag 
level with other extant calibrations of the targets presented here, except
in the $I$ band, where there are color dependent deviations of up to 0.05 
mag versus one other photometric sequence. There is no 
clear resolution of this difference: we present as much verification of 
the sequences presented here as possible. We argue that a very likely 
reason for such discrepancies is differences in the filter bandpass.
\end{abstract}


\section{Introduction}

Accurate absolute photometric measurements of stars and galaxies lie at
the very heart of modern astronomy. Modern day measurements, both from space 
with the \textit{Hubble Space Telescope}, as well as with large aperture 
telescopes from the ground are routinely being used for projects that require 
accurate photometry.  Examples range 
from the distance scale based on Cepheid or RR Lyrae variables
or supernovae, to metallicity estimates from the position of the red giant
branch, to the ages of globular clusters and dwarf galaxies throughout the
Galactic halo. With each improvement in accuracy it becomes possible to
attack whole new classes of problems.

It is fundamental to precision measurement, that calibrators and science
targets be observed in exactly the same way to minimize systematic
errors.  This has not been possible for most photometric observations
using the \textit{Hubble Space Telescope} (\textit{HST}) or large ground-based telescopes,
due to the lack of faint standards. For instance, the photometric
zeropoints for the Wide Field Planetary Camera~2 (WFPC2) on board \textit{HST} have been
monitored using exposures of a few seconds of a $V = 13$ star, while 
most science observations utilize exposures of typically 500--2000$\,$s,  
and are interested in much fainter targets (e.g., Cepheids from $V$ = 24 to 26).  

In 1994, Peter Stetson (see \citet{Kelson96} and \citet{Saha96}) 
first noticed that his
WFPC2 photometry from short and long exposures of Palomar~4 and NGC~2419
did not agree at the few percent level.  This led to a more complete analysis
by \citet{CasMut98}\footnote{see 
http://www.stsci.edu/instruments/wfpc2/Wfpc2\_isr/wfpc2\_isr9802}, and by
\citet{Stetson98} which confirmed that the WFPC2 has
a non-linearity problem, which is responsible for absolute
photometric uncertainty that could be as large as 0.05 mag.
Thus far, zero-points reported by various investigators scatter with rms 
values like $0.02$ mag for $V$ and $I$, and $0.03$ to $0.04$ mag for 
$B$ and $U$ \citep{Heyer04}\footnote{see http://www.stsci.edu/instruments/wfpc2/Wfpc2\_isr/wfpc2\_isr0401.html}. 
A brief history of attempts to understand and mitigate the anomalous behavior of WFPC2 can be found in 
\citet{Dolph00b}.
 
The obvious solution is to obtain fainter standards for use with \textit{HST}
and large ground-based telescopes which have similar problems.  In a
recent compilation, \citet{Stetson00} presented $> 15,000$ new
photometric standards in {\it BVRI\/}.  While Stetson's current 
database\footnote{http://cadcwww.dao.nrc.ca/cadcbin/wdb/astrocat/stetson/query}
does contain samples of standard stars in a number of globular clusters and 
dwarf galaxies that have been observed over the 
past decade with WFPC2, it is still highly desirable to increase the number 
density on the sky, the magnitude limit, and the accuracy of faint standard 
stars in a number of historically popular science fields for two reasons:
\begin{enumerate}
\item
Fields suitable for direct 
observation with \textit{HST} instruments must have sufficient number of stars per 
square arc-minute to efficiently map the Quantum Efficiency (QE) and 
Charge Transfer Efficiency (CTE) of all the component CCD chips.   
\item
Observing established standards {\it now} with WFPC2
would not tell us anything about the historical calibration 
of the instrument, since it has been shown \citep{Whitmore99} that the 
calibration of WFPC2 for faint objects has evolved with time. Rather, 
fields that have been observed with WFPC2 at multiple epochs during its 
active life should be calibrated as best as possible, so that the 
time varying behavior of the instrument can be characerized after the fact.
\end{enumerate} 

We have therefore used WIYN, with its relatively large
aperture and excellent seeing, to extend the Landolt \textit{BVRI} photometric system
\citep{Landolt92, Landolt83}  
to fields that have already been repeatedly observed with \textit{HST} and which
are known to have a suitable density and magnitude range of stars to
provide an effective retroactive calibration of WFPC2. 
We are able to obtain consistent photometry to levels of 
$0.01$ mag down to $V \approx 22$ mag in our targets. 
In this paper we describe our experiment and 
present the results of this calibration for three such 
suitable targets: the globular clusters NGC~2419, Pal~4, and Pal~14.
We demonstrate that systematic and random errors have been contained 
sufficiently well so that stars in these fields have been tied to the Landolt
system with accuracy $ \sim 0.01$ mag or better. 

The results from this project should improve the magnitude gap 
between photometric standards and science targets. For instance, typical 
corrections for WFPC2 linearity will be over a
dynamic range of $\approx$ 10 (between photometric standards and
extragalactic Cepheids, for instance) rather than the prior range of
$\approx$ 100,000. These calibrated fields will also help calibrate new 
generations of \textit{HST} imagers. Our photometry of NGC~2419 has already been 
used to verify the calibration for the {\it Advanced Camera for Surveys (ACS)} 
\citep{Sirianni} and will provide an efficient way to perform nightly photometric 
calibrations for imagers on 8-m class telescopes.

In the following sections we describe the observations used, followed by a 
discussion of the peculiarities of the instrument configurations used, and the 
methods required to properly process the photometry extraction that 
successfully removes the instrumental signatures. The photometry for 
four target fields (NGC~2419, Pal~14, Pal~4 and Pal~3) are presented.
The exposure to exposure and night to night consistency is examined in 
a way that gives separate estimates for systematic and random errors in 
calibration. Lists of stars in each of the four fields are presented, with the 
measured magnitudes and error estimates. Comments on how they may 
be best utilized as standard stars is discussed. 

\section{Instrumental Setup}

Observations were made primarily with the WIYN 3.5-meter telescope at the 
Kitt Peak National Observatory. The Mini-mosaic ({\it MIMO}) camera mounted 
on one of the Nasmyth focii was used. This instrument has two $4096 \times 
2048$ {\it SITE} CCD chips, mounted side by side, providing a roughly square 
field of view (FOV). The 15 micron pixels project to $\sim 0.14$ arc-seconds 
on the sky at the $f/6.4$ Nasmyth focus of the telescope. This is an excellent 
match to the superb image quality delivered by the telescope (median $\textit{FWHM} 
= 0.7$ arc-sec), and also provides an FOV of about 9.6 arc-minutes on a side.
The imaging performance of this telescope and instrument combination is 
described in \citet{Saha00}. Each chip is read out by two amplifiers as two
$4096 \times 1024$ sub-sections. 

The instrument was commissioned with rigorous tests which cover linearity, 
shutter timing performance, charge transfer efficiency, dependence of 
accumulated counts from a star with background level, and geometrical 
distortion of the focal plane. The complete retinue of tests is described 
in the commissioning report, which can be found at the website:
{\it http://www.noao.edu/wiyn}. The charge transfer inefficiency is too 
small to be noticed, even at the lowest background levels. Shutter timing 
and/or shading corrections are unnecessary at the 0.2\% level, 
even for exposures as short as 0.3s. No measurable dependence of aperture 
photometry on position in the field of view (due to geometric distortion, 
for instance) could be detected. With the exception of a few issues that we 
mention below, the photometric performance of {\it MIMO} is better than 
$0.2$\%. Throughout the course of the observations reported in this 
paper, one of us (AS) continually monitored the performance of the instrument, 
to ensure that it remained satisfactory. 

The only real issue of note has been the existence of a cross-talk between 
amplifiers: if a pixel is saturated in the A/D counter of one amplifier, the 
corresponding pixel in the sub-section read by each of the other amplifiers can 
be contaminated. This problem is described in the commissioning report, and 
methods to mitigate it in the data reduction process are provided in an anonymous 
ftp area that can be reached from the {\it Documentation/} tag on the website:
{\it http://www.noao.edu/wiyn} under the sub-heading {\it Saha's notes on 
reducing MINIMO data}. Pixels that are affected by cross-talk were masked, 
and not used in subsequent analysis. 

The spectral response of the chips from 380 nm to 850 nm varies smoothly 
from about 80\% in the middle of this range to no worse than 50\% at the ends.
The response drops sharply from 850 to 950 nm, a spectral region that is 
within the $I$-band. The two chips are very similar in their spectral response, 
although that does not alleviate the need to evaluate color terms separately 
for each. The filters used (\textit{BVRI}) are the so called `Harris set', which 
closely resemble the canonical passbands. The transmission data for the 
actual filters used are available on the website: {\it http://www.wiyn.org/filters}, 
where they are designated {\it W2, W3, W4}, and {\it W5} ({\it B, V, R} and {\it I} 
respectively). The point to note is that the filters are well bounded, and have no 
red-leaks even beyond 1000~nm.  Photometry of standard stars shows that the 
overall spectral response (telescope plus filters plus detector) requires only 
linear color terms with coefficients smaller than $0.05$. 

There are a few items of note regarding the telescope and instrument
combination. Since this is an alt-az telescope, the pupil rotates with respect 
to the image plane, and any azimuthal asymmetry in pupil illumination will 
require flat-field correction to be a function of the image rotator position.
Dome flats were obtained covering many orientations of the instrument base. 
The worst peak to valley differences in flat-field images taken in different 
image rotater positions was $0.5$\%. The next issue is that the image port 
was designed to include a field-flattener, which is also an atmospheric 
dispersion compensator. Unfortunately this optical component produces 
considerable light loss, especially in the blue, and so is not used. Over the 
FOV of {\it MIMO}, the lack of the field flattener can produce small but
noticeable changes in the shape of the PSF, especially when the seeing 
is good. This must be accounted for when doing PSF photometry.
It appears that there is also a `vertical' displacement in the relative 
mounting of the two chips of perhaps a few microns, which is sufficient to 
produce a subtle discontinuous change in the PSF shape as one steps from 
one chip to the other.  In total, the PSF changes have been confirmed to 
be a function only of the radial position in the field (as one would expect if 
the source of the problem is the absence of the field-flattener) plus a step 
from one chip to the next. The resulting complications for PSF fitting 
photometry are adequately handled if aperture corrections are determined as 
a function of radial position on {\it MIMO} in addition to a step function going 
from one chip to the other. The third item of note is the imperfect baffling 
at the Nasmyth port, which allows some stray scattered light to enter the camera. 
We have ascertained that this produces only an elevation of the background 
(`sky') level, and changes over spatial scales that are much much larger than a 
stellar PSF. As a result there is no contribution due to this on aperture 
or PSF photometry of stars.

It is thus clear that there are no issues with this setup that would preclude 
obtaining photometry with systematic errors much smaller than one percent. 
The telescope-instrument-site combination has delivered $R$~band images with 
{\it FWHM} as small as 0.28 arc-sec (without any adaptive optics corrections)
which is as well as one can expect to do from the ground. Median delivered 
image quality is between 0.6 and 0.7 arc-sec. In order to calibrate faint sequences 
within small fields of view, which is necessary for the retro-active calibration of the 
\textit{HST} image archives as well as being best suited for work with current 8-m 
class telescopes, optimal image quality is essential to minimize confusion noise.  
The {\it WIYN/MIMO} combination is thus as good as one might expect to have. 
The only net disadvantage is the slow read-out time of the chips, which limits the 
efficiency when observing the bright Landolt standards. It would have been better 
to observe more standard stars than we were able to do, but on pristine 
photometric nights the self consistency in the photometry of standard stars was 
seen to be more than satisfactory, leaving us confident about the results.

In addition, our target fields and standard stars were also observed at the WIYN 
0.9-m telescope at Kitt Peak, using S2KB, a $2048 \times 2048$ SITE chip.  
Photometry of our target fields with this independent setup (telescope, detector, 
filters, are all different) is restricted to brighter magnitudes, but is adequate for 
revealing any systematic errors that arise from either telescope, i.e.\ concordant 
photometric results is a strong argument for absence of systematic errors 
in both setups. 

\subsection{Observations}

Observations of NGC~2419, Pal~4, and Pal~14, all of which are distant globular 
clusters in our galaxy and have the additional virtue of having being observed 
repeatedly with the {\it WFPC2} camera of the \textit{HST}, were made on nights 
that appeared to be cloudless. Contemporaneous observations of Landolt 
equatorial standards were also made. Pointings that include several Landolt 
stars within the FOV of {\it MIMO} were chosen, taking care also to choose 
fields that contain standard stars with a wide range of colors. Exposure times 
on the Landolt fields were set to optimize the signal-to-noise (S/N) ratios for 
the standard stars. Varied exposure times were used on the target fields, 
which allows the calibration of stars over a wider range of magnitudes, while 
also providing a running check on the linearity of the system (i.e.\ by 
comparing the measured magnitudes of stars that appear bright on long 
exposures, but faint on short ones). On some nights images were taken 
so that different exposures were taken with two camera base orientations set 
$180^{o}$ from one another. This not only puts each star once on each chip, 
but also reverses the direction of charge transfer with respect to the field, 
thus providing closure tests not only on chip-to-chip response variations, 
but also on losses from charge transfer problems. Standard star fields 
were observed over a wide range of airmasses, so that target field 
observations were well bracketed. Exposure times for standard stars were 
always at least one second.

The true test of whether a given night was photometric was decided after 
reducing the standard star data. On good nights, typical rms residuals in the 
photometry for standard stars per measurement are $\sim 0.01$ mag or 
smaller: errors in the mean are typically $\sim 0.003$ mag.  Only nights 
which proved to be photometrically pristine were used in the final analysis.  
A list of these nights, with a summary of the observations made, is given in 
table~\ref{T1}. The processed images from all the nights are being made 
available in the STScI archive.

\section{Data Processing and Photometry}

The raw images were pre-processed to identify and mask out known cosmetic 
flaws on the detectors, and saturated pixels and pixels affected by the  cross-talk 
described in \S2. A program written in IDL by one of us (AS) was used. Bias 
subtraction for the {\it MIMO} CCDs is best done on a line by line basis, which 
was also done as part of the pre-processing.  Flat-field corrections were done 
in {\it IRAF}, using the external {\it mscred} package. 

\subsection{Instrumental magnitudes of the Standard Stars}

Aperture photometry of the Landolt stars were done interactively, using an 
IDL program written by AS. A measuring aperture of 35 pixel radius was used, 
which corresponds to an aperture of $\sim 5$ arc-sec radius. This is a large 
aperture in pixels, as a result of the fine spatial sampling of 0.14 arc-sec per 
pixel, and so the standard stars need to be well exposed in order to avoid 
read-noise of the CCD (or shot noise from the sky) from contributing uncertainty.  
An initial background value was estimated from an annular aperture from 40 to 
60 pixels in radius, and the value of the sky was interactively adjusted so that an 
aperture growth curve is flat near 35 pixel radius. This procedure works well for 
images with seeing $\textit{FWHM} \leq 1.8$ arc-sec, which sets the limit for the 
worst delivered image quality (DIQ) that can be tolerated. There is some low 
level light in the PSF even beyond the 35 pixel aperture boundary, but as long 
as the $\textit{FWHM} \leq 1.8$ arc-sec, it is entirely due to scattered light in 
the optics (mostly from the tertiary mirror), which does not change from one 
exposure to the next in the same passband. In other words, the 35 pixel aperture 
measures the same fraction of light from a star in different exposures with 
different \textit{FWHM}, as long as the $\textit{FWHM} \leq 1.8$ arc-sec. 

\subsection{PSF Photometry and Instrumental Magnitudes of the Target Stars}

PSF photometry was run on the target fields with a modified version of the 
{\it DoPHOT} program \citep{Schechter93}. For a given image, the PSF is not 
allowed to vary with position in the field of view. The private version of 
{\it DoPHOT} that was used, produces two sets of aperture magnitudes 
(for a range of measuring aperture radii) of the brighter relatively uncrowded 
stars {\it in isolation} (i.e.\ with all other objects subtracted after fitting). One 
set is for background value chosen so that the growth curve is flat near 15 
pixel radius, and the second set is for background value chosen so that it is 
flat near 35 pixels. Let us designate the instrumental aperture mag at 15 pixel
radius from the first set by $m(15)$, and that at 35 pixel radius from the 
second set by $m(35)$. The latter is equivalent to the interactively measured 
instrumental aperture magnitude described above for the standard stars. The 
advantage here is that we measure the star in isolation (with all neighbors 
subtracted), and do so automatically for all the stars (from several to several 
hundreds, depending on the target field), that have adequate S/N. The 
disadvantage is that an automatic algorithm for flat growth curve background
determination can go awry from specious or unsubtracted features in the 
image.  However, this problem is mitigated by the number statistics from a 
large number of stars.

If $m(fit)$ is the PSF fitted magnitude, one needs simply to find the 
dependence of $apcorr_{35} = m(35) - m(fit)$ with detector position $(x,y)$ 
to account for the subtle spatial variation of the PSF. This is not easy in 
practice, since there are not always a sufficient number of stars in the FOV 
for which $m(35)$ can be measured with sufficient S/N. However, since the 
PSF variation is primarily due to subtle focus changes and lack of perfect 
field flattening over the field area, only the core of the PSF varies with 
position on the FOV, while the wings of the PSF remain unchanged. 
Experimentation shows that the PSF variation due to position in the FOV 
is essentially all contained within a 15 pixel radius of the center of the stellar 
profile. Thus the spatial variation of the PSF can be compensated by 
calculating $apcorr(15) = m(15) - m(fit)$, and mapping it as a function of 
position in the FOV. There are more stars for which one can obtain $m(15)$ 
with high enough S/N than there are for $m(35)$.

Using even a few stars for which both $m(15)$ and $m(35)$ are measured 
with high S/N, we can derive the additive offset required to go from 
a 15 pixel radius aperture magnitude to a 35 pixel radius aperture mag, since 
we only need one quantity, which is invariant over the FOV. Thus, 
\begin{equation}
  \textit{offset}_{15}^{35} = \langle m(35) - m(15) \rangle
\end{equation}
Meanwhile, $apcorr_{15} = m(15) - m(fit)$ is fitted as a function of 
position $x$ and $y$ on the FOV:
\begin{equation} 
  apcorr_{15} = apcorr_{15}(x,y)
\end{equation}
With experimentation, we have found that a quadratic polynomial with circular 
symmetry about the optical axis (center of the FOV) is adequate (a general 
two dimensional quadratic function has the risk of being unconstrained in the 
corners if there are not enough stars to define it), plus an additional term to 
correct the discontinuity going from one chip to the next, as described in \S2.  
The sign and magnitude of the variation depends on the seeing, as well as 
on the focus setting (i.e.\ how far, and in which direction, the telescope is 
from true focus). Once the polynomial description of $apcorr_{15}$ (in 
terms of $x$ and $y$), and the value of $\textit{offset}_{15}^{35}$ are known, 
the PSF fitted magnitude $m(fit)$ for any star anywhere in the FOV can be 
put on the system of 35 pixel apertures, which we designate as the 
instrumental magnitude $ m^{instr}$. 
\begin{equation}
  m^{instr} = m(fit) + apcorr_{15} + \textit{offset}_{15}^{35}
\end{equation}
This brings the target star instrumental magnitudes to a common footing 
with the instrumental magnitudes of the standard stars.

Software necessary to determine and apply the aperture corrections as above 
were custom written in the {\it IDL} language.

\subsection{Nightly Photometric Solutions}

To account for extinction due to airmass, and to allow for color-terms and 
zero-point adjustment in order to transform to $m^{true}$, the true magnitude 
on the Landolt system, we can write a system of equations of the kind:
\begin{equation}
 m_{i}^{true} = m_{i}^{instr} ~+~ C_{i} ~+~ \alpha_{i} X ~+~ \beta_{i,j} (m_{i}^{instr} - m_{j}^{instr})
\end{equation}
where $i$ and $j$ are indices for the various passbands ($i \neq j$), and $X$
is the airmass at which the object is observed. The standard star observations 
are used to solve for the coefficients $C, \alpha$, and $\beta$.\footnote{
Note that in this formulation, one can additionally use the instrumental mags 
of any stars in the object fields that were observed more than once, and at 
significantly different air-masses to assist in the determination of $\alpha$.}
$C_{i}$ and $\beta_{i,j}$ are in general different for each chip, and were 
solved for separately for each chip.  Higher order color terms might be 
required, but with our setup, as described in \S2, were found to be 
unnecessary. Also, no cross-terms of airmass with color were used. Such 
terms are sometimes necessary, since selective absorption by the 
atmosphere can alter the effective central wavelength for a passband, i.e.\ 
a star (particularly if it is of extreme color) observed at high airmass is seen 
with a different effective wavelength than if observed at low airmass. 
All of our target observations were made at airmass less than $1.7$, and
observations of Landolt stars were done up to airmasses of 2.0. 
The use of such cross-terms did not improve the nightly solutions: typical
star by star rms scatter in the standard star solutions was $\sim 0.015$ mag 
or smaller, with no discernible trends in the residuals with the product of 
airmass and color. 

In the following set of equations, the solution for the night of 2003 Feb~9 (UT) 
for one of the CCD chips (using 22 measurements in each passband) is 
shown, illustrating typical values for the coefficients. The lower case symbols 
$b,v,r,i$ denote instrumental magnitudes in units of $-2.5 \log (ADU s^{-1}) + 
30.00$, whereas $B,V,R,I$ denote magnitudes on the Landolt system. To 
derive the coefficients in the equations below, Landolt's published photometry 
\citep{Landolt92} was used to derive the constants and coefficients in the 
equations below (and for each night of observations in this program).  The 
errors in the mean are shown in parentheses. Not all color combinations are 
displayed.\footnote{While it is common practice to express the night constants 
and color coefficients with the true magnitudes and colors as the independent 
variables, and instrumental magnitudes as the dependent terms, we have 
chosen to do the opposite, as in the above, for ease of computing the true 
magnitudes for the target stars, and because this formulation can be used to 
estimate atmospheric extinction by comparing the instrumental magnitudes of 
stars in the target fields when they are observed at different airmasses. This 
enables the inclusion of the lower end of airmass values not available from 
Landolt's equatorial standard stars when observing from mid-latitudes.  In the 
absence of non-linear color terms, the two forms are algebraically identical. 
Further, the errors in measurement, i.e.\ of the instrumental magnitudes, are 
insignificant, and the direction of the regression makes no difference. 
Nevertheless, we have verified this assertion by computing the regression the 
other way, and find there is no difference at the 0.001 mag level in how these 
equations are specified and calculated.}
\begin{eqnarray}
   B &= &b - 4.120 - 0.262 X + 0.013 (b-r)  ~~~(\pm 0.003) \\
   V &= &v - 4.095 - 0.146 X - 0.010 (v-i)  ~~~(\pm 0.002) \\
   R &= &r - 3.978 - 0.103 X - 0.012 (b-r)  ~~~(\pm 0.002) \\ 
   I &= &i - 4.714 - 0.065 X - 0.024 (v-i)  ~~~~(\pm 0.003)
\end{eqnarray}

Over the course of the several nights of observations, we noted quite significant 
changes in the extinction coefficients. The pattern in the night to night 
differences is that a roughly equal  value is added to the extinction coefficients 
of {\it all} passbands. For instance, on the night of 2003 June~4 (UT), the 
extinction coefficients as above for $B,V,R,I$ were $0.350, 0.221, 0.169$, and 
$0.124$ respectively.  The extinction in all bands is uniformly increased by $0.07 
\pm .02$ relative to 2003 Feb~9. It is possible that dust, water-vapor and aerosol 
play a role in this, and it is wise, therefore, to restrict observations to relatively low 
airmass, since the scale height of these contaminants may be different from 
other atmospheric sources of scattering.

Once the photometry solution for a night is obtained, as above, the instrumental 
magnitudes $m^{instr}$ of each star on a set of target observations (obtained 
from aperture correction of the DoPHOT PSF fitting magnitudes according to 
the procedure described in \S4.2) in at least 2 passbands can be converted to 
the Landolt system. DoPHOT reports good estimates of the (random) 
measurement errors from the photon statistics, read-noise characteristics and 
fit residuals for each star. These were propagated along with the calibrated 
magnitudes for the next steps in analysis. 

\section{Analysis of Errors}\label{ERRS}

The procedures described above yield magnitudes that are nominally calibrated 
on the Landolt system for each set of exposures (in the various passbands). 
We see that the nightly photometric solutions themselves carry errors that are 
smaller than one percent, which gives us the expectation that the systematic 
errors are also adequately small. However, it is necessary to see how well the 
systematic errors are indeed contained, by comparing the results obtained 
across several nights of observations, particularly since extinction coefficients 
have been seen to vary significantly from night to night. We have data from 
several nights for each of our targets, including some observations made 
with an independent setup (0.9m telescope and a different physical set of 
filters). We also sometimes have several determinations for magnitudes from 
different exposure sets on the same night. However, there is no external 
estimate of how well the aperture correction procedures worked, and 
systematic errors from that are not separately accounted for. In the final 
analysis, the repeatability of the magnitudes from all of these different 
measurements is the true test of how well systematic errors have been 
contained. Also the final reported magnitudes for each star should be a 
mean that is adjusted and weighted appropriately from all the photometric 
epochs at which observations were made.

We describe here the adopted procedure, using as an example, the case of 
$V$ magnitudes for the NGC~2419 field. There are 14 epochs over 7 nights. 
After cross-matching the stars from all the epochs, an initial weighted mean 
value $\overline{V_{k}}$ for the $k^{th}$ star was calculated using the inverse 
of the variance of individual magnitude errors reported by DoPHOT as weights. 
The stars that are common to all 14 epochs were then identified: in the case at 
hand, 74 such stars were found.  If $V_{j}^{i}$ denotes the $i^{th}$ 
measurement for the $j^{th}$ such common star, and $e_{j}^{i}$ is the 
corresponding measurement error estimate reported by DoPHOT, the quantity
\begin{equation}
\Delta V^{i} = \frac{ \sum_{j} w_{j}^{i} (V_{j}^{i} - \overline{V_{j}})  }{ \sum_{j} w_{j}^{i} }
\end{equation}
where
\begin{equation}
\label{dophweights}
w_{j}^{i} = \frac{ 1 }{ (e_{j}^{i})^{2} }
\end{equation}
is the best estimate of the systematic zero-point offset for magnitudes measured 
in the $i^{th}$ epoch, relative to the mean of all epochs. More specifically, the 
error in the mean from the rms scatter of the $\Delta V^{i}$'s is a measure of the 
systematic error in the values of $\overline{V_{k}}$, and we denote it by 
$\epsilon_{sys}$.\footnote{We adopt the convention of reporting rms scatter 
with $\sigma$, `errors in the mean' with $\epsilon$, and measurement
uncertainties with~$e$.}
\begin{equation}
\label{syserr}
(\epsilon_{sys})^{2} = \frac{ \sum_{i} (\Delta V^{i})^{2} }{ N(N-1) }
\end{equation}

$\Delta V^{i} $ was then subtracted from the magnitude of every star as measured 
in the $i^{th}$ epoch (for each of the epochs), i.e.:
\begin{equation}
{\cal V}_{j}^{i} = V_{j}^{i} - \Delta V^{i}
\end{equation}
from which we recompute the final weighted mean magnitude of the $k^{th}$ star as:
\begin{equation}
\label{Vmagwt} 
\overline{\cal V}_{k} = \frac{ \sum_{i} w_{k}^{i} {\cal V}_{k}^{i} }{ \sum_{i} w_{k}^{i} }
\end{equation}
where the $w$'s are the same as in equation \ref{dophweights}. We also obtain the 
reduced chi-square for each star as:
\begin{equation}
\label{chisq}
\chi_{\nu,k}^{2} = \frac{ \sum_{i} w_{k}^{i} (({\cal V}_{k}^{i} - \overline{\cal V}_{k})^{2}  }{ \nu - 1 }
\end{equation}
where $\nu$ is the number of available observations for star $k$. 

The weighted random error in the mean for the $k^{th}$ star, $ \epsilon_{k} $, is then given by:
\begin{equation}
\label{Vwterr}
(\epsilon_{k})^{2} =  C. \frac{ 1. }{ \sum_{i} w_{k}^{i} }
\end{equation}
where $C$ is set to unity if $\chi_{\nu,k}^{2} \leq 1$, and $C = \chi_{\nu,k}^{2}$ otherwise.  
This formalism deals optimally with the fact that the various exposures are of different 
depth, and thus a given star in general has very different signal-to-noise from one 
exposure to the next. Further, $\epsilon_{k}$'s are reliable estimates of the uncertainty in 
the individual $\overline{\cal V}_{k}$'s, irrespective of whether the DoPHOT reported error 
estimates $e_{k}^{i}$'s are correct. In the case of a variable object, the value of $\epsilon$ 
is the uncertainty in the mean brightness. 

In reporting our results, we will keep the systematic errors $\epsilon_{sys}$ separate from 
the random/variability uncertainties $\epsilon$. The latter are reported star by star, whereas 
the former are the same for all stars for any target field and filter. Table~\ref{T2} gives the 
values of $\epsilon_{sys}$ for each target and passband. Random errors will be reported 
in later sections regarding the individual targets.

\section{Results}

For each of 3 targets for which we have sufficient data (namely NGC~2419, 
Pal~14 and Pal~4) we present here the following data products:
\begin{enumerate}

\item A deep reference image of the target field is shown both in print, as
well as a \textit{FITS} file in the electronic version of this paper. The \textit{FITS} 
files are appointed with a `world coordinate system') (WCS) calibrated 
to J2000 coordinates on the sky.  All positions of individual stars for these 
target fields are referred to positions ($X$ and $Y$ in pixels) on the 
corresponding image and/or their J2000 positions. 

\item For each target, a table of the stars for which there are at least 3 
measurements in each of the four passbands $B, V, R,$ and $I$, and where 
$\epsilon_{k}$ (as in equation~\ref{Vwterr}) is less than 0.015 mag in each 
passband. These are the best stars to be used as standards. This table is 
given both in printed form, and also in the electronic version of this paper. 
This table is numbered for easy reference to individual stars. In using these 
values, one should be mindful also of the additional systematic errors tabulated 
for each passband and target field in Table~\ref{T2}.

\item A table of all available objects with measured magnitudes in at 
least $V$ and $I$ is presented as an electronic table.

\item A table identifying objects flagged as variable stars, where the selection 
criterion is that the star have $\chi_{\nu,k}^{2} $ (as in equation~\ref{chisq}) 
higher than 20 in all of $B, V, R,$ and $I$, with at least 6 measurements in 
each passband. Also, the rms scatter $\epsilon_{k})$ is required to be greater 
than 0.05 mag in $V, R,$ and $I$, and greater than 0.1 mag in $B$. This table 
is also numbered for easy reference to individual stars. 

In the following sub-sections, we identify and comment on these data products
target by target.
\end{enumerate}

\subsection{NGC 2419}

\begin{figure}
\epsscale{.4}
\plotone{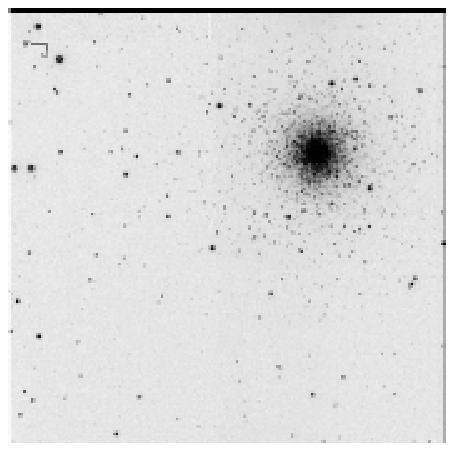}
\caption{Chart of the target field for NGC~2419. North is to the left and East is down. A \textit{FITS} file of this 
$4150 \times 4100$ pixel image is given in the electronic edition of the {\it PASP} and at ftp://taurus.tuc.noao.edu/pub/saha/Photseq/fg1\_elec.fits. The positions in pixels in the FITS image correspond to the $X$ and $Y$ positions of  given in the tables for NGC~2419 objects in this paper. The FITS image is also appointed with a world coordinate system (WCS) so that sky coordinates can be read directly from the image (using a suitable display program). 
\label{n2419chart}}
\end{figure}

The reference field for NGC~2419 is shown in Fig.~\ref{n2419chart}. The 
\textit{FITS} image supplied in the electronic version reveals much more detail, 
depth, and dynamic range, and shows individual stars much further into the 
cluster core than the printed version.  

Table~\ref{N2419best} gives the magnitudes and uncertainties (from random 
errors) in four passbands for all the stars in the field (designated `best' stars) 
for which the random uncertainty in the mean is less than 0.015 mag in all the 
passbands.  The corresponding data for all stars for which $V$ and $I$ 
magnitudes are available are given in Table~\ref{N2419VIexist}.
Finally, the stars which are most likely variables (at least 6 available 
measurements in each of the four passbands, and with $\chi^{2}_{\nu} > 20$ 
in each passband) are listed in Table~\ref{N2419variab}.

\begin{figure}
\epsscale{0.4}
\plotone{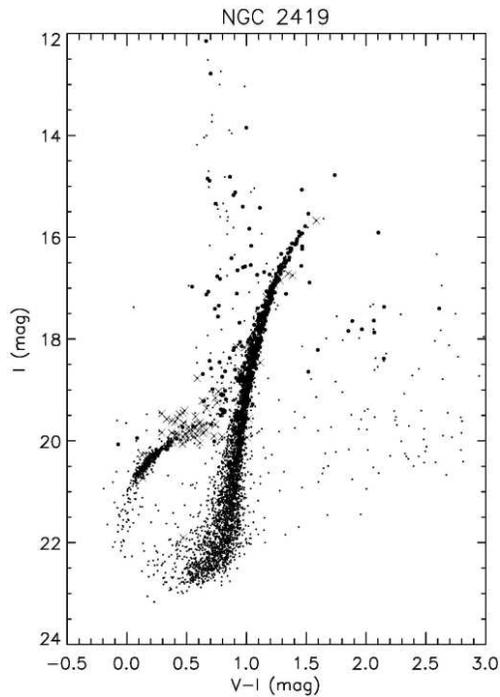}
\caption{The color-magnitude diagram in $V$ and $I$ for the NGC~2419 field. 
Bold points mark objects with photometric uncertainty better than 0.015 mag in 
all passbands. Crosses mark variable stars. Note that the `best' stars
span a range of over 8 magnitudes, and a color range of over 3 mags in $V-I$.
\label{N2419cmd}}
\end{figure}

A color-magnitude diagram (CMD) in $V$ and $I$ for stars in the NGC~2419 
field (utilizing all the objects listed in Table~\ref{N2419VIexist}) is shown in 
Fig.~\ref{N2419cmd}. The `best' stars are marked in bold, and the variables 
are also identified. In addition to cluster stars, there are clearly several 
foreground stars, including a few `best' stars with colors redder than $V-I \sim  
1.5$ mag. There are also `best' stars among the blue horizontal branch stars in 
the cluster to colors as blue as $V-I \sim 0.0$. While the `best' stars go no
fainter than $I \sim 20.5$ mag, taking ensemble averages for the large number
of fainter stars still permits one to directly calibrate photometry with 
uncertainties from random errors contained below 0.01 mag. 

The systematic error estimates (given in Table~\ref{T2}) in $V, R$ and $I$ for 
NGC~2419 are below 0.006 mag rms. This includes 3 nights of observations 
with the independent setup at the WIYN 0.9-m telescope, and these small 
values thus strengthen our confidence that the magnitudes given in the tables 
in this paper are truly on the Landolt system. The systematic error estimates 
must be added in quadrature to the individual rms estimates for random errors.
The larger systematic error of $ \sim 0.02 $ mag in $B$ is disappointing. The 
seeing was often poorer in $B$, and bright uncontaminated stars with sufficient 
S/N that are common to exposures that range in depth by a factor of 100, and 
observed in common on two telescopes of very different aperture, were few. 
This has resulted in poor S/N for anchoring the $B$ photometry to the Landolt scale. 

A total of 92 objects were deemed to be variables. Fig.~\ref{N2419cmd} shows 
that the majority of these are in the location of the RR~Lyrae region in the 
CMD, with a few that are brighter than the horizontal branch by $\sim 0.5$ mag.
A few lie along a suggestive track that leads up along where the AGB should
be, with a few more yet up among the brighter red giants. There is a solitary 
variable among where one find SX~Phe or W~UMa stars. No attempt was made to 
cross identify with known variables.

\subsection{Pal~4}

\begin{figure}
\epsscale{.4}
\plotone{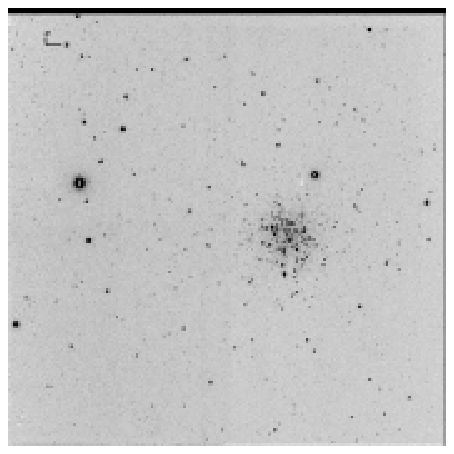}
\caption{Chart of the target field for Pal~4. North is to the right and East is up. A \textit{FITS} file of this  
$4150 \times 4100$ pixel image is given in the electronic edition of the {\it PASP}  and at ftp://taurus.tuc.noao.edu/pub/saha/Photseq/fg3\_elec.fits. The positions in pixels in the FITS image correspond to the $X$ and $Y$ positions of given in the tables for Pal~4 objects in this paper. The FITS image is also appointed with a world coordinate system (WCS) so that sky coordinates can be read directly from the image (using a suitable display program). 
\label{Pal4chart}}
\end{figure}

The reference field for Pal~4 is shown in Fig.~\ref{Pal4chart}. As before, the
\textit{FITS} image supplied in the electronic version reveals much more detail, 
depth, and dynamic range. 

Table~\ref{Pal4best} gives the magnitudes and uncertainties from random sources 
for the `best' stars in Pal~4, i.e.\ those for which the uncertainties are less than 
$0.015$ mag in each of the four passbands. Table~\ref{Pal4VIexist} presents the 
magnitudes for all stars in the field for which at least $V$ and $I$ magnitudes are 
available. The likely variables are identified in Table~\ref{Pal4variab}. The systematic 
uncertainties for each passband listed in Table~\ref{T2} for Pal~4 must be added 
(in quadrature) in all cases to get the true uncertainties with respect to the Landolt system.

\begin{figure}
\epsscale{0.4}
\plotone{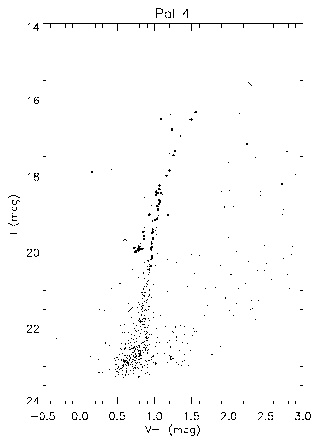}
\caption{The color-magnitude diagram in $V$ and $I$ for the Pal~4 field. 
Bold points mark objects with photometric uncertainty better than 0.015 mag in 
all passbands. Crosses mark variable stars. Note that the `best' stars
span a range of nearly 5 magnitudes. 
\label{Pal4cmd}}
\end{figure}

The CMD for the target field is given in Fig~\ref{Pal4cmd}. It is much sparser than 
for NGC~2419: even the numbers of foreground stars is fewer for this field, which is 
less than 20 degrees from the North Galactic Pole. Also, the data from the 0.9m 
telescope did not have enough stars with high enough S/N at levels faint enough for a 
good overlap with the 3.5m data. As a result, there are no calibrated stars in this field 
that are brighter than $\sim 16$ mag. There is only one `best' quality star bluer than 
$V-I = 0.5$ and only two that are redder than $V-I = 1.5$. However, there are several 
stars in the giant branch and red clump which can be used as photometric standards to 
$I \sim 20$ mag, and fainter calibration can be had from the ensemble average 
of stars to $I \sim 22$ mag.

Three variable stars are identified: from their location on the CMD, one is likely an 
RR~Lyrae star, another an SX~Phoenicis star, and the brightest object is either an AGB 
variable (if it is a cluster member---it does lie in the line of sight of the cluster) or a field 
variable of as yet unknown type.

\subsection{Pal~14}

\begin{figure}
\epsscale{.4}
\plotone{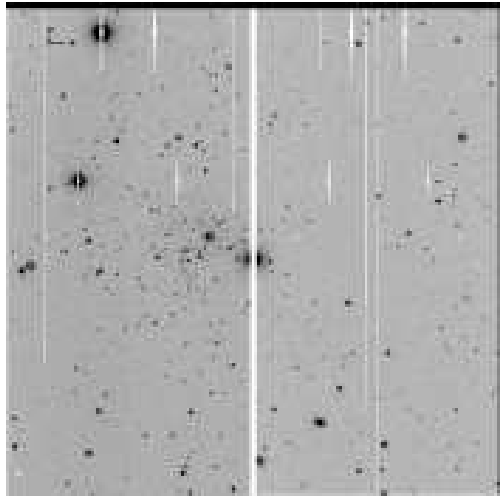}
\caption{Chart of the target field for Pal~14. North is to the right and East is up. Due to the desire to position the bright star near the cluster
between the two chips of MIMO, no dithering was done in the `X' direction, and so this stacked  deep retains the masked areas. A \textit{FITS} file of this $4150 \times 4100$ pixel image is given in the electronic edition of the \textit{PASP} and at ftp://taurus.tuc.noao.edu/pub/saha/Photseq/fg5\_elec.fits. The positions in pixels in the FITS image correspond to the $X$ and $Y$ positions of given in the tables for Pal~14 objects in this paper. The FITS image is also appointed with a world coordinate system (WCS) so that sky coordinates can be read directly from the image (using a suitable display program). 
\label{Pal14chart}}
\end{figure}

The reference field for Pal~14 is shown in Fig.~\ref{Pal14chart}. Once again, 
the \textit{FITS} image supplied in the electronic version reveals much more detail, 
depth, and dynamic range.

Table~\ref{Pal14best} presents the photometry for the `best' stars, which have 
uncertainties less than $0.015$ mag in each passband. Table~\ref{Pal14VIexist} 
gives the available magnitudes for all stars for which $V$ and $I$ magnitudes 
could be measured, and Table~\ref{Pal14variab} lists the likely variables. As 
before, systematic errors with respect to the Landolt system as listed in 
Table~\ref{T2} must be added in quadrature to all listed uncertainties. 

\begin{figure}
\epsscale{0.35}
\plotone{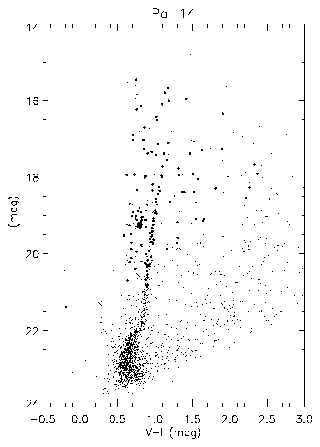}
\caption{The color-magnitude diagram in $V$ and $I$ for the Pal~14 field. 
Bold points mark objects with photometric uncertainty better than 0.015 mag in 
all passbands. Crosses mark variable stars. Note that the `best' stars
span a range of over 5 magnitudes. 
\label{Pal14cmd}}
\end{figure}

The CMD is given in Fig.~\ref{Pal14cmd}. The field surrounding the cluster is more 
populous than for Pal~4, and the `best' stars span the range of colors better. 
Unfortunately again, as in the case of Pal~4, the 0.9m observations do not produce 
enough stars with sufficient overlap to allow the calibration of standard stars brighter 
than $I \sim 15.5$ mag. At the faint end, the sub-giant branch is well defined by 
numerous stars, and (as in the case for NGC~2419 and Pal~4) even though they do 
not individually qualify in the `best' category, ensemble averages of several of these 
stars can carry the calibration well beyond $I \sim 21$ mag. 

Seven variable stars are identified, some of which are likely to be foreground field stars.

\section{Comparison with other sequences}
\label{Intercompare}

\subsection{Comparison with Stetson's sequence in NGC~2419}

\citet{Stetson00} has presented a photometric sequence in NGC~2419 that 
spans a similar field size and brightness range. A comparison with this 
independent calibration is instructive. For NGC~2419, there is also an older 
unpublished sequence in \textit{UBVRI} by L.~Davis in the KPNO consortium 
fields \citet{ChristianXX}. Comparison with that is also of interest.

\begin{figure}
\epsscale{0.8}
\plotone{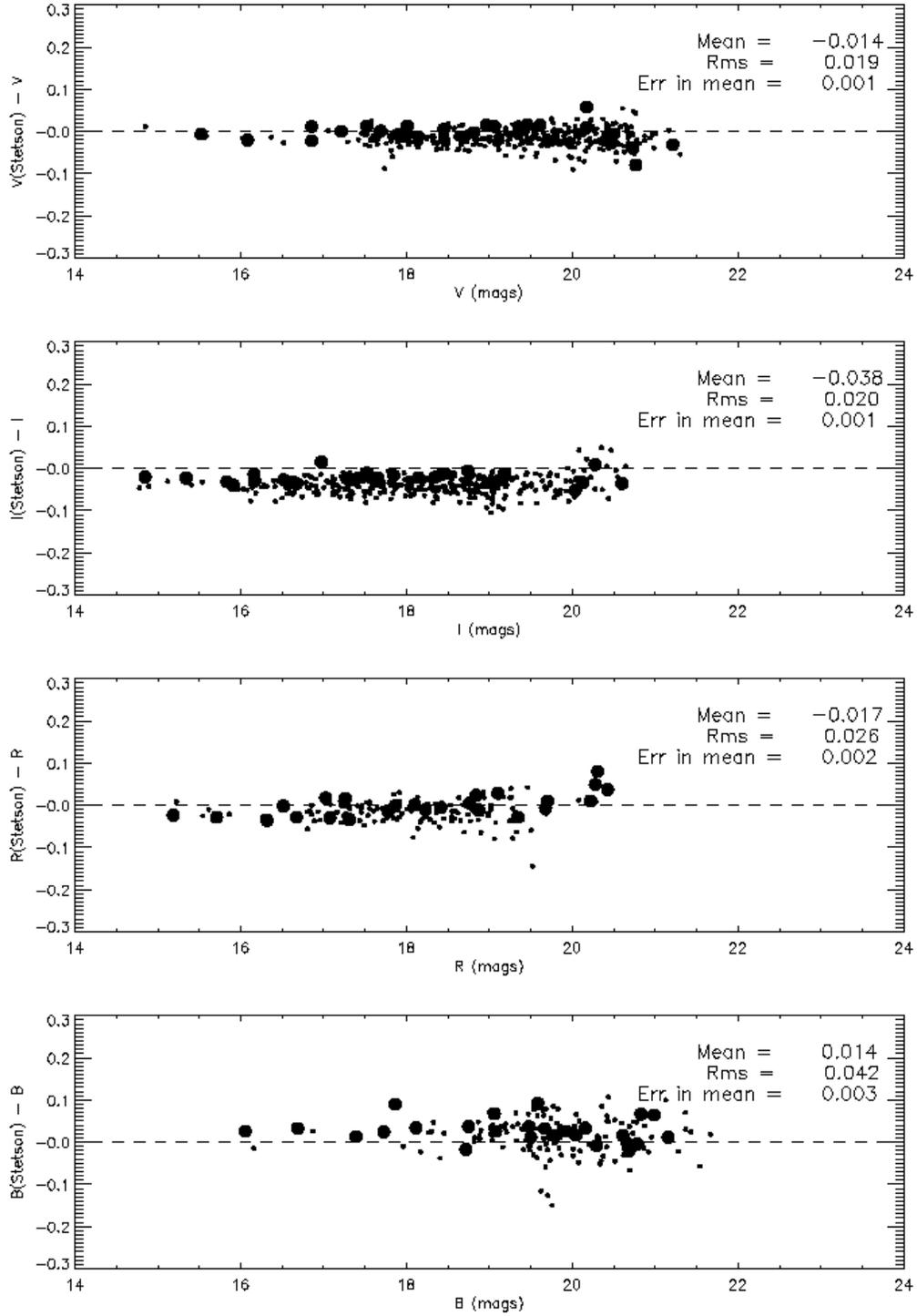}
\caption{The star by star differences (shown as small dots) in all four 
passbands of the magnitudes of the `best' stars from this study versus the 
magnitudes reported in \citet{Stetson00}. The differences are shown as a 
function of object brightness. No obvious trends with brightness are seen.
The unweighted mean differences 
with uncertainties are shown in the annotations.
The large dots show the differences versus the \citet{ChristianXX} sequence. 
\label{Stetcompa}}
\end{figure}

The comparison in all four passbands with the \citet{Stetson00} sequence 
for NGC~2419 is shown in Fig.~\ref{Stetcompa}: the small dots show the 
difference in magnitudes star by star versus the mean magnitude of the 
objects as reported in this paper. The annotations in the figure give the net 
differences (unweighted means for all objects, and the errors in the mean) 
in the sense of Stetson minus this work. The mean differences are larger 
than can be accounted by the sum (in quadrature) of the error in the mean 
and the systematic error estimates in Table~\ref{T2} (except in the $B$ 
band where the systematic error estimate is atypically high). Thus, they 
cannot be ignored, and must be discussed. In $V$ and $R$ bands, the 
Stetson sequence is on average brighter by $\sim 0.015$ mag, but in the 
$I$ band it brighter by almost $0.04$ mag. In comparison, the 0.9m and 
3.5m observations reported in this paper when taken separately,  differ 
(in the sense of 0.9m minus 3.5m) by $-0.005 \pm .006$, $+0.020 \pm 0.004$, 
and $ +.024 \pm 0.010$ mag in $V, R$, and $I$ respectively, which are smaller
than the difference between the Stetson sequence and this work. Note also, 
that there is no obvious brightness dependence in the difference between the 
Stetson scale and the work in this paper: non-linearity is not the culprit.  

The large dots in Fig.~\ref{Stetcompa} show the difference between the 
L.~Davis unpublished sequence \citep{ChristianXX} and the present work. On 
average the Davis sequence appears to lie midway between the Stetson and 
current work values. We should note that formally the Davis sequence is tied 
to \citet{Landolt83}, not \citet{Landolt92}, though differences due to this are 
not expected to be significant beyond a few milli-mags.

\begin{figure}
\epsscale{0.8}
\plotone{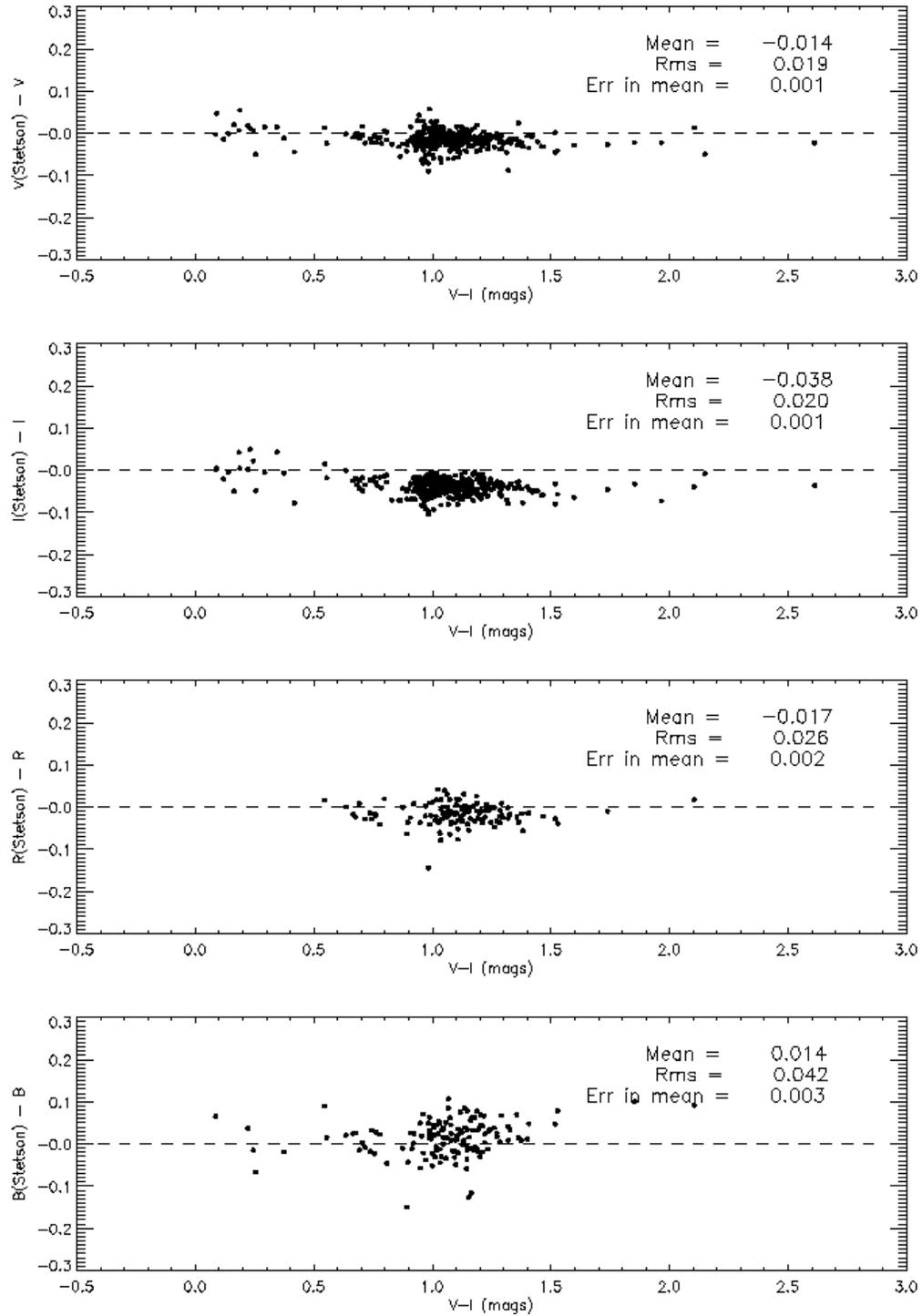}
\caption{The star by star differences in all four 
passbands of the magnitudes of the `best' stars from this study versus the 
magnitudes reported in \citet{Stetson00}. The differences are shown as 
a function of object color. Clear trends are seen in $V$ and $I$, especially 
in $I$, indicating that the source of the differences is likely in how well 
the two observation sets duplicate the original passband.
\label{Stetcompb}}
\end{figure}

An important characteristic of the difference between the \citet{Stetson00}
sequence (especially in $I$) and the work in this paper is revealed in 
Fig.~\ref{Stetcompb}, where the differences between the two sequences is 
shown star by star as a function of color ($V-I$).  Note how both in $V$ and 
especially in $I$, the differences are a function of color. Specifically at $V-I 
\sim 0.2$ there is no difference on average in either $V$ or $I$, whereas at 
$V-I \sim 1.2$, The Stetson sequence is brighter in $I$ by $\sim 0.05$ mag, 
and in $V$ by $\sim 0.02$ mag. There are not enough data in $R$ to see if 
a color dependence exists, and in $B$ the differences are within the systematic 
error estimate (which is regrettably large).

\subsection{Comparison with other photometry in NGC~2419}

\begin{figure}
\epsscale{0.8}
\plotone{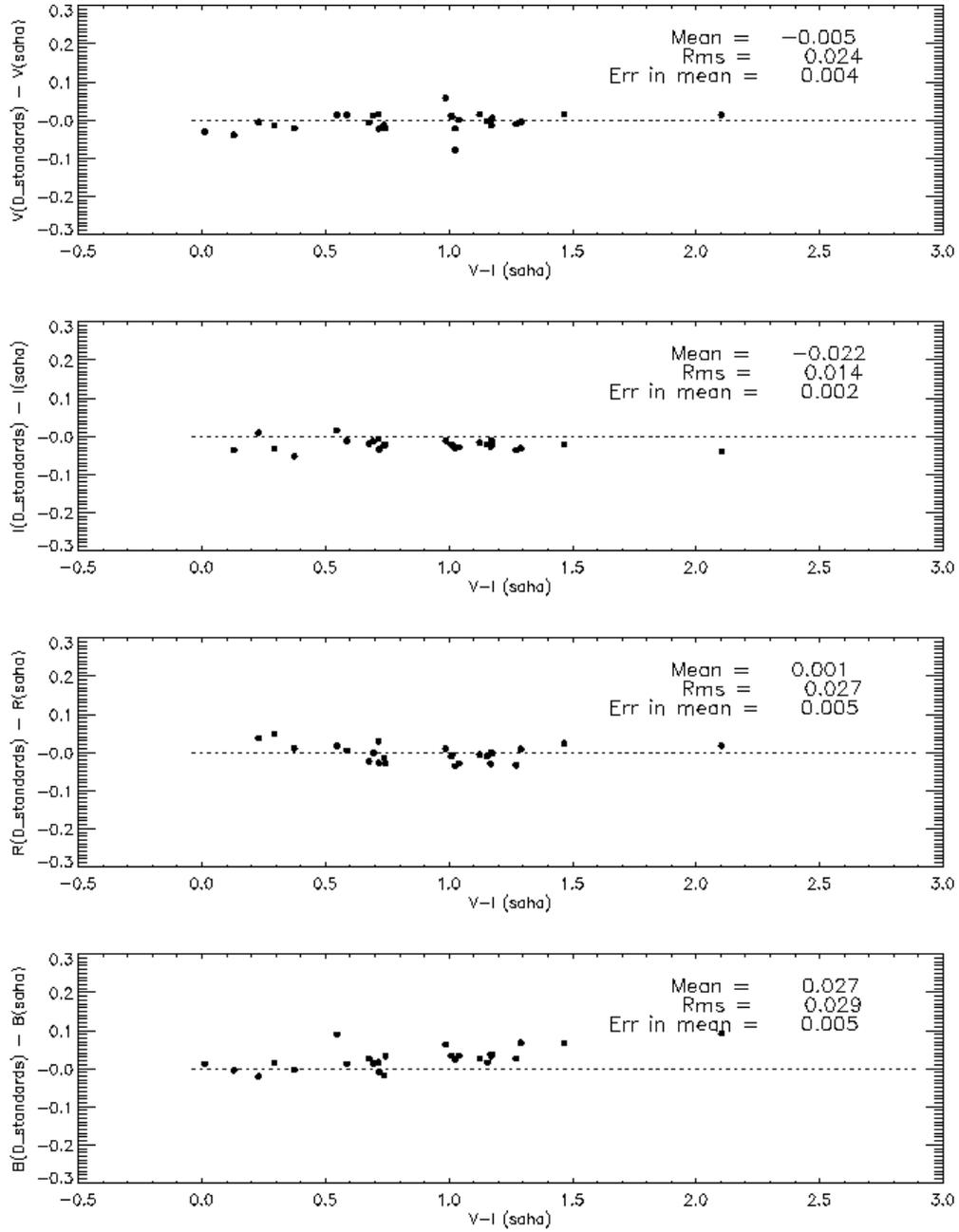}
\caption{Same as Fig~\ref{Stetcompb}, but the comparison is against the 
unpublished sequence by Davis for the Kitt Peak consortium (see text).
The trends seen for $V$ and $I$ in Fig.~\ref{Stetcompb} are not seen here, 
but a trend appears for the $B$ band. 
\label{Dcompb} }
\end{figure}

The equivalent comparison of magnitude difference as a function of color against 
the Davis standards in NGC~2419 is shown in Fig.~\ref{Dcompb}. The annotations 
show the net differences and level of significance similar to those given for the Stetson 
sequence in Fig~\ref{Stetcompa}. The color dependent trends seen against the 
Stetson sequence are not present here.

\begin{figure}
\epsscale{0.8}
\plotone{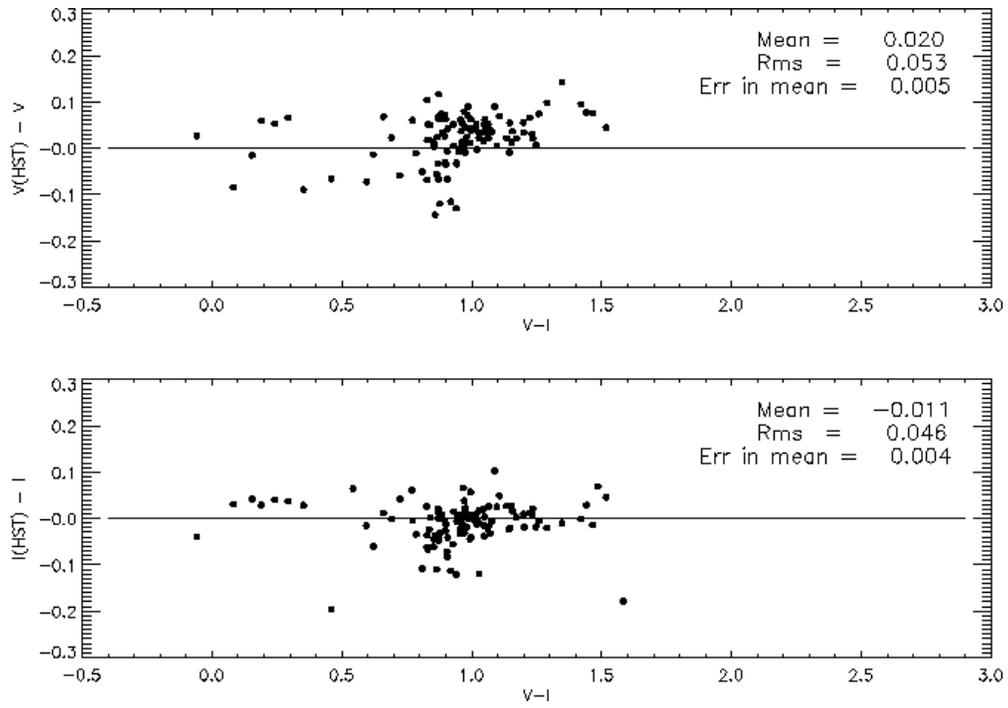}
\caption{Comparison of $V$ and $I$ magnitudes against photometry 
with WFPC2 of stars in an area just south of NGC~2419. The \textit{HST}/WFPC2
data were reduced with HSTPHOT (see text for details) using independent 
zero-point and color terms. The color trends seen in Fig~\ref{Stetcompb}
are not discernible here.
\label{HSTcompa} }
\end{figure}

One of us (AED) has obtained photometry in $V$ and $I$ of stars with WFPC2 in a 
field south of NGC~2419, which overlaps in area with our ground based field here. 
The photometry were obtained using the {\it HSTPHOT\/} program described in 
\citet{Dolph00a}, using a procedure that corrects for the CTE anomalies described 
in \citet{Dolph00b}. Independent photometry zero-points and color-terms derived 
from photometry in $\omega$ Cen by \citet{Walker94} were used.  The comparison 
of these magnitudes with those in this paper is shown in Fig~\ref{HSTcompa}. The 
data span a smaller range in color than in Fig~\ref{Stetcompb}, and the blue stars 
show significant scatter (because the blue horizontal branch stars are faint on these 
relatively shallow WFPC2 exposures).  Despite these shortcomings, the two sets 
of photometry show better agreement than the comparison with the Stetson 
NGC~2419 sequence. Specifically, if the color trend seen against the Stetson 
sequence were present, it would be revealed by these data, but no such comparably 
strong trend is seen. 

We have private communication from Sirianni that the synthetic photometry calibration 
of the ACS on \textit{HST} is in close agreement with the results presented in this paper.

The various comparisons detailed above are summarized in Table~\ref{Comptable}.

We see that there is still a lack of concordance between different independent 
investigations regarding the calibration of the $I$ band at the few percent level. 
Systematic differences of a few hundredths of a magnitude, especially in the 
$I$ band, are apparently not uncommon. Many $I$ filters in use at various 
observatories allow out of band transmission in the infrared. While detectors like 
the S20 extended response photomultipliers were blind at near infra-red 
wavelengths, CCDs can have substantial response at $1.0$ to $1.2~\mu$. 
At these wavelengths, the spectra of cool stars have pronounced bands, and 
simple color transformations (even with higher order terms) may not adequately 
account for out of band transmission, since color `excesses' become 
discontinuous. In addition, as we show in the Appendix, the form of the standard 
color equations used in practice is a poor match to physical reality unless the 
color response mis-match between the employed and the original measuring 
systems are small, i.e.\ correctable by a term that depends only on the first 
moment in frequency/wavelength of the spectral energy distribution.

Whether the above is in fact the reason for the difference between the Stetson 
sequence and the one derived here, is not established beyond doubt. In practice 
we see much smaller deviations because of bandpass mis-match, which arises 
out of the fact that the spectral energy distribution of stars vary only in highly 
constrained ways. Synthetic simulations with a possible extreme variant of the $I$ 
band (see \S\ref{Banddiff}) shows that at most half of the discrepancy seen can 
be explained this way. The idea of using the 0.9m data as an arbiter, while sound 
in principle, is only marginally useful because of the lack of S/N at faint magnitudes. 
While the comparison through the HSTPHOT photometry of WFPC2 observations 
in the NGC~2419 field against the system of \citet{Walker94} is consistent with the 
sequence derived in this paper, it is at least a little bit circular, given that we are 
trying to set up a sequence to retro-actively calibrate WFPC2. The agreement
of the NGC~2419 photometry presented here against the ACS photometry by 
\citet{Sirianni} (priv.\ comm.) as discussed above, is an endorsement of the 
sequence derived here. However the agreement may be viewed with some reserve 
since the ACS Photometry results are from a synthetic calibration. 

\section{Verification of our own photometry \label{Verif}}

\begin{figure}
\epsscale{0.8}
\plotone{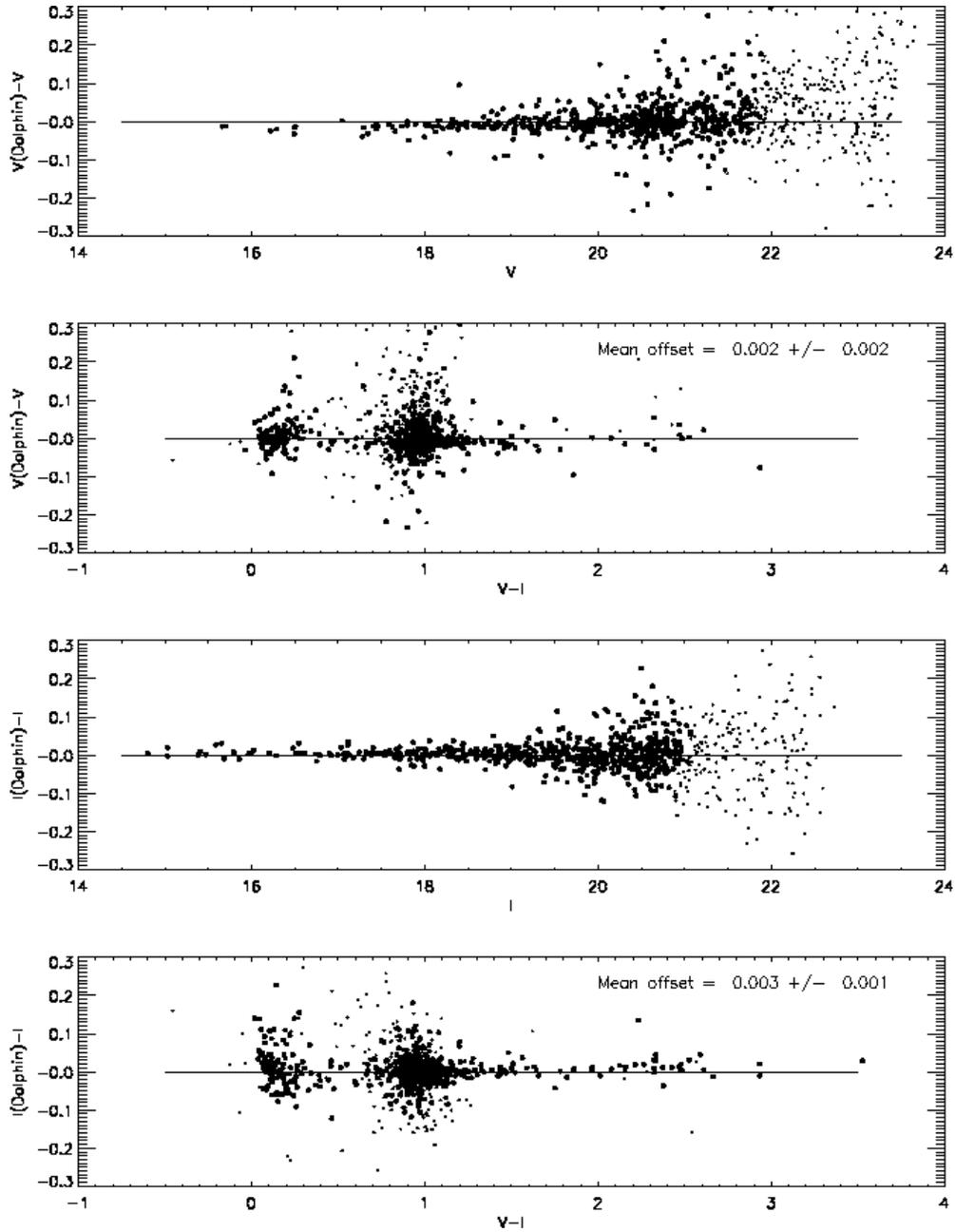}
\caption{Comparison of $V$ and $I$ magnitudes from two different 
independent reduction methods applied to images obtained on 2003~Feb~09.
$V$ and $I$ refer to magnitudes obtained with the methods described in 
this paper, whereas $V(Dolphin)$ and $I(Dolphin)$ are magnitudes derived 
using independent methods of PSF fitting, atmospheric extinction estimation, and color 
equation determination by one of us (AED). The excellent agreement 
corroborates that systematic errors arising from differences in methodology 
are insignificant.
\label{DolphcompareVI} }
\end{figure}

As a sanity check to test that there is not some hidden error in our reduction, one 
of us (AED) has independently reduced observations on the night of 2003 Feb~09. 
All measurements were made independently, including PSF fitting, which was done 
using a modified version of HSTPHOT \citep{Dolph00a} and solving for extinction 
and color terms. The results from identical data frames were compared. The 
comparison in $V$ and $I$ bands are shown in Fig~\ref{DolphcompareVI}. The 
ensemble mean differences in photometry, derived from stars that are reported 
individually (by the respective reduction procedures) to have measurement 
errors less than 0.05 mag, is only a few milli-mags for both passbands. No 
color dependent trends are visible.

\subsection{Testing the Assumptions in the Calibration Procedure}

Our discrepancy with Stetson's sequence begs further introspection. It can be 
asked whether inclusion of a quadratic term in the color equation for the $I$ 
band color-equation would bring the results from our data into better agreement 
with Stetson sequence. In other words, have we neglected a color term that 
needed to be included? It can be alleged, that our sparse observations of Landolt 
standards (due to instrument read-out time limitations) does not permit a quadratic 
term to be well constrained from  a single night's worth of data. To refute this 
argument, we demonstrate below that while our observations are much less 
extensive than Stetson's, it is adequate for asserting that the discrepancy seen is 
a real disagreement, and not an artifact of inadequate data.

The first argument is that $\epsilon_{sys}$ in Table~\ref{T2} value for $I$ in NGC~2419 
is $.0058$ mag. This value comes from 14 measurements on 7 different nights on 2 
different set-ups. The extinction coefficients and color equations were derived 
independently for each night using only observations from the same night. Despite 
noting quite large changes in extinction from one night to another, and observing on two 
different set-ups, the value of $\epsilon_{sys}$ is satisfactorily small. Since this is an 
`external' estimate of the systematic error, it is very unlikely that systematic errors in 
calibration have been made at the $0.04$ mag level.

\begin{figure}
\epsscale{0.60}
\plotone{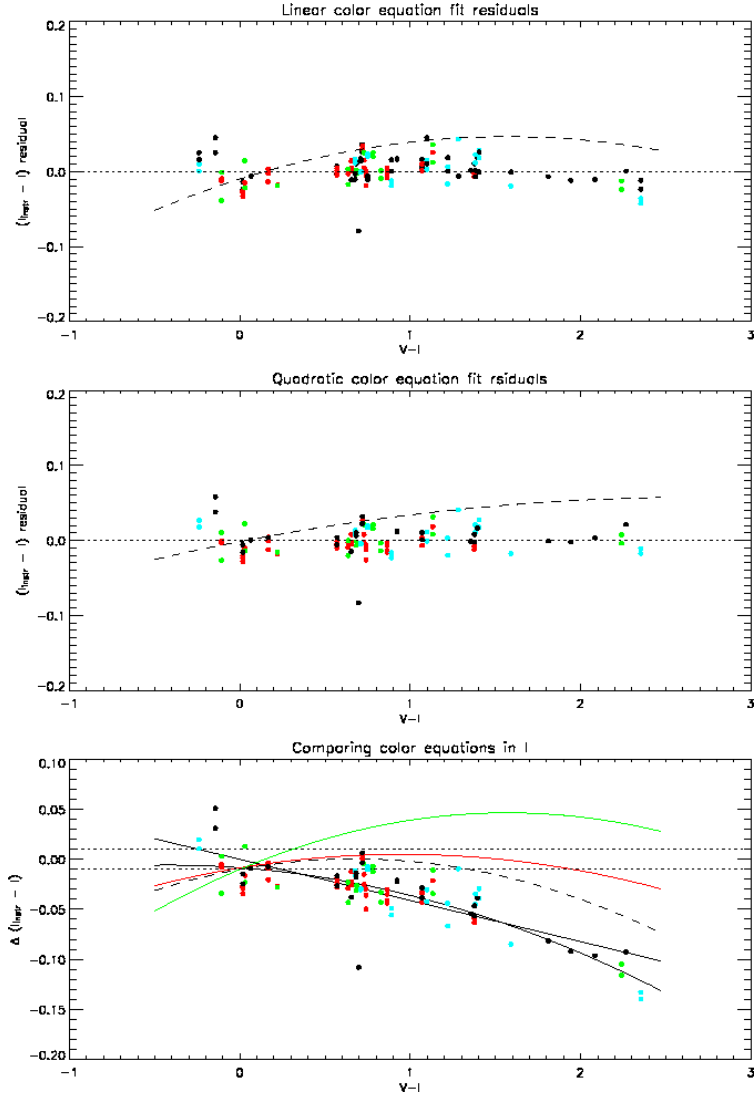}
\caption{Residuals from simultaneously fitting the color equations to 
standard star data from 4 different nights in the manner described 
in \S\ref{Verif} are shown in the first two panels: the top panel shows 
the case for a linear fit, and the second panel for a quadratic fit. The 
various colors mark data from different nights. The dashed line in the first
two panels show how much the fit residuals would have to change 
systematically with color to produce concordance with Stetson's sequence in 
NGC~2419. The scatter in the standard star data is sufficiently small to rule 
out the possibility that the discrepancy is due to uncertainties in our 
determination of the color correction. In the lowest panel, we show the 
color equations themselves: the straight and curved continuous lines are our 
best linear and quadratic solutions respectively; the dashed line is the color 
equation required to bring our $I$ band observations in  NGC~2419 into 
agreement with Stetson's sequence; the red line shows the difference between 
our quadratic and linear fits and shows that for $ -0.1 < V-I < 2.0$ the 
difference is less than 0.01 mag, and the green line shows the difference 
required from our linear fit to match Stetson's NGC~2419 sequence. More 
details are in \S\ref{Verif}. 
\label{verify} }
\end{figure}

The second argument involves a combined analysis of standard star observations 
made on 4 different nights when $V$ and $I$ observations of NGC~2419 were made
with the WIYN 3.5-m telescope set-up. In this check, we assume for simplicity 
that the color equations are the same for both CCD chips (their spectral responses 
are very close), and that they do not change from one night to another. Allowing for a 
different zero-point for each night, we wish to derive the color coefficients. We use 
the extinction coefficients for $I$ as originally derived: recall that our original procedure 
corrects the instrumental magnitudes, and the atmospheric coefficients are derived 
not just from the standards, but from all high S/N stars (including those in our target 
fields) that were observed at different airmasses. This correction reduces the 
instrumental magnitudes to zero air-mass. An error as large as $0.04$ in the 
extinction coefficient for $I$ will produce relative errors between standard stars 
observed at the extremes of airmass range (1.15 to 1.95) of $0.03$ mag. The rms 
scatter (per star) in determining the extinctions is seen to be at least a factor of 3 
smaller, and errors in the mean from all the measured standards are thus a few 
milli-mags at most. Thus extinction determination cannot itself be a significant 
source of systematic error. Let $i$ denote the instrumental magnitudes corrected 
to zero air-mass.  We then write the usual expression:
\begin{equation}
\label{linfit}
i - I = L_{0} + L_{1} (V - I)
\end{equation}
which is the linear form of the color equation, and where $V$ and $I$ are Landolt's 
values for the standard stars. According to the discussion above, $L_{0}$ can change 
from night to night, but $L_{1}$ is assumed the same for all observations of standards.
Similarly, one can write the quadratic form as:
\begin{equation}
\label{quadfit}
i - I = Q_{0} + Q_{1} (V-I) + Q_{2} (V-I)^{2}
\end{equation}
We can solve these equations for the standard stars, allowing only one value each for 
$L_{1}$ and for $Q_{1}$ and $Q_{2}$, but allowing $L_{0}$ and $Q_{0}$ to differ from 
night to night. The residuals for the fit to equation~\ref{linfit}are shown in the top panel of 
Fig.~\ref{verify}, with data points from the four different nights coded in four different 
colors. The middle panel shows the residuals for the same observations for the fit to 
equation~\ref{quadfit}. In each of the above panels, the dashed lines shows the locus 
about which the points would lie had we fitted the instrumental mags of stars in 
NGC~2419 to the Stetson sequence.  These figures demonstrate that the difference 
seen with respect to the Stetson sequence is real, and not the figment of inadequate 
data. Nor is the difference attributable to using a linear color term when a higher order 
term is demanded.  The continuous black lines in the third panel of Fig.~\ref{verify}
show the fits to equations~\ref{linfit} and \ref{quadfit} (with respective $L_{0}$'s and 
$Q_{0}$'s subtracted). The dashed shows the equivalent quadratic fit, if instead of 
fitting Landolt standards, the instrumental mags of stars in NGC~2419 are fitted to 
the Stetson sequence. The red line shows the difference purchased by fitting a 
quadratic versus a linear color equation to the Landolt stars alone: note that in the 
range $-0.1 < V-I < 2.0$, the difference is always smaller than 0.01 mag. Contrast this 
with the green curve, which shows the difference between fitting a quadratic color 
equation to the Stetson sequence in NGC~2419 versus a linear fit to Landolt standards. 
Since the fit to Stetson's sequence produces a discrepancy with the Landolt standards 
at a level of significance that is clear from the 3 panels of Fig.~\ref{verify}, the 
discrepancy cannot be an artifact of inadequate observations of standards. It is a real 
discrepancy, clearly evident in the data, even though it lacks a clear physical explanation 
(except possibly the discussion in the Appendix).

\subsection{Is the subset of Landolt standards actually used skewed from the 
Overall Landolt System?}

Another possibility is that the subset of Landolt standard stars used in our calibration 
is somehow skewed from the parent set of all Landolt standards. In particular, 
Landolt's photo-electric measurements were made with a 14 arc-sec aperture, 
whereas here we used 10 arc-sec (diameter). In the present work we preferred those 
fields where a number of standard stars are present within the instrument's field of 
view, so that we could maximize the number of standards observed while keeping the 
number of exposures (and accompanying large overhead in read-out time) 
manageably small. Thus we preferred those fields where the chance of having another 
star within a 14 arc-sec aperture is increased relative to fields with relatively fewer 
available stars. We made all our standard star measurements interactively, and any 
gross cases of such object confusion were immediately apparent, and the offending 
object was not used further. However, no explicit procedure to guard against such an 
occurrence was used in a systematic manner, and it is possible, though unlikely, that 
systematic differences between Landolt's measurements and ours have been 
introduced in this manner.  It is further unlikely that such an error has occurred only in 
one of the four bands, and even more so that the difference is correlated with color. 

\begin{figure}
\epsscale{0.75}
\plotone{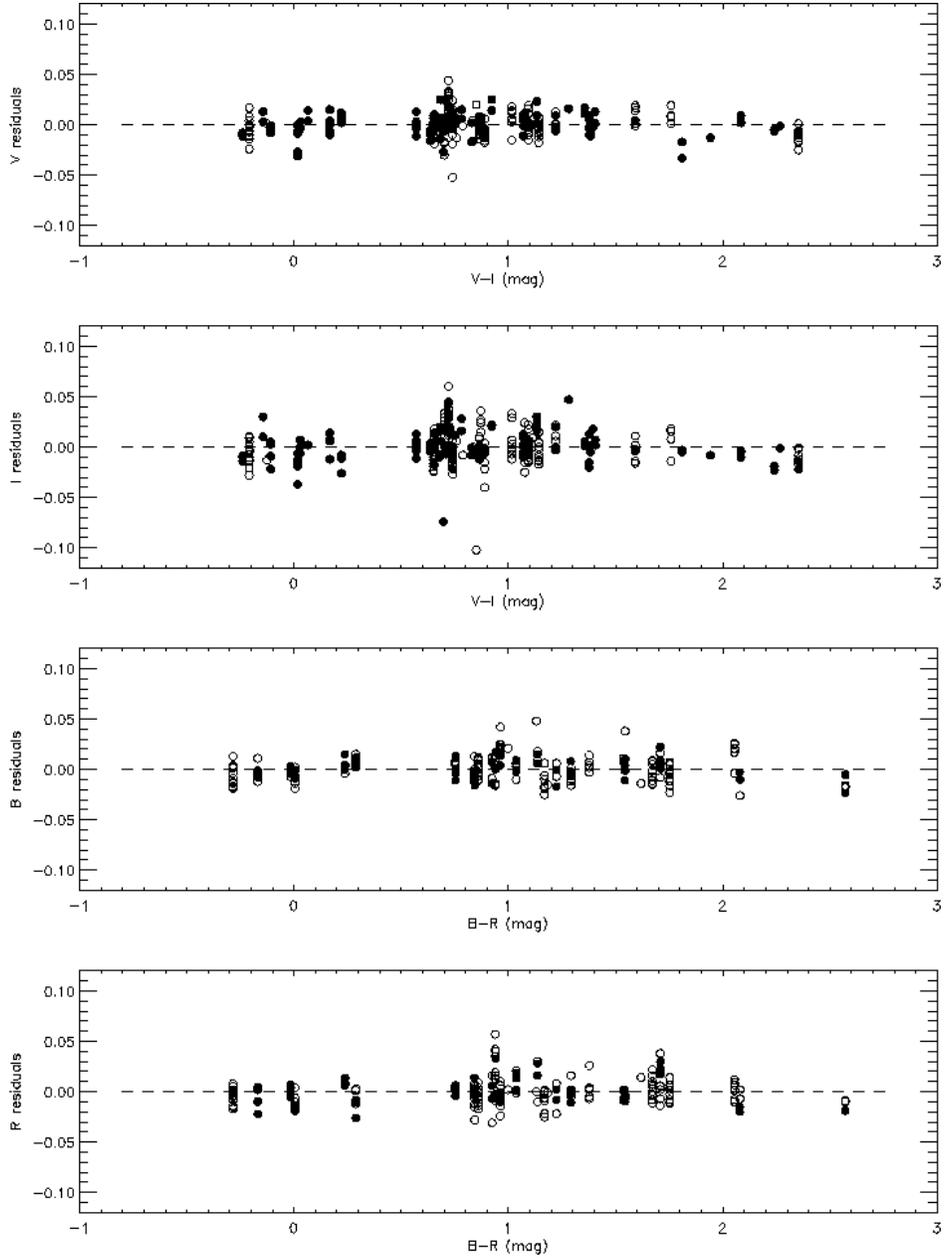}
\caption{Residuals in all 4 passbands for the Landolt stars observed on all 
photometric nights are shown. The residuals are with respect to the 
predicted value using the photometric solution for the night in question. 
Filled circles show those observations that contribute specifically 
to the calibration of NGC~2419, while open circles contributed only to 
Pal~4 and/or Pal~14.
\label{Stdres} }
\end{figure}

Fig.~\ref{Stdres} shows the results of the photometry of Landolt stars.  Each point on 
the ordinate for each pass band is the residual (observed minus calculated, after fitting 
the photometric solution for the relevant night) for each standard star measurement 
(one point per star per observation). The filled circles are for observations on nights 
that have contributed to the calibration of NGC~2419. This is like the first panel of 
Fig.~\ref{verify}, but for each of the four bands, and using the color terms evaluated 
independently for each night. No perceptible systematic difference is seen for the 
observations relevant to NGC~2419 when compared to the rest. 

\begin{figure}
\epsscale{0.75}
\plotone{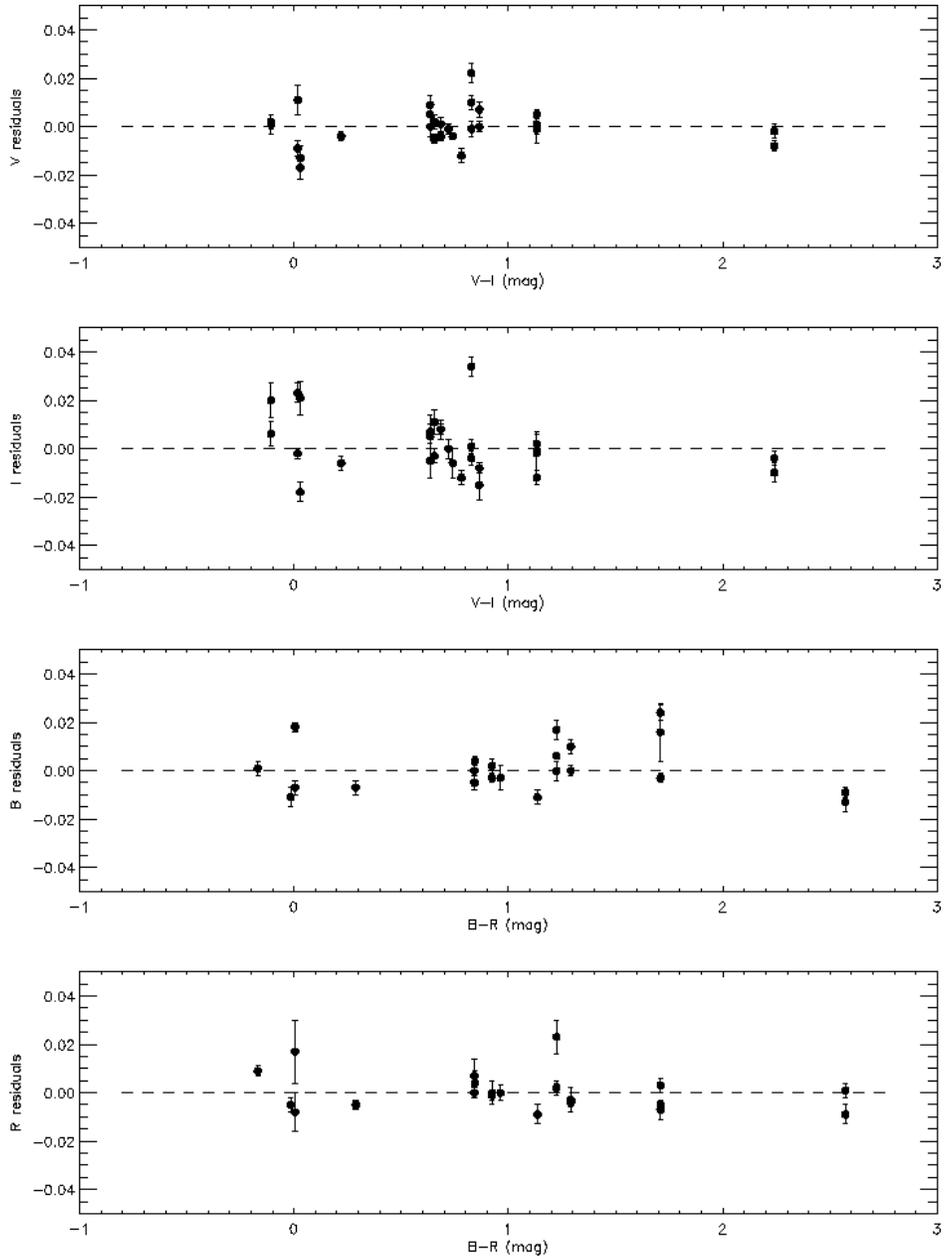}
\caption{Same as Fig.~\ref{Stdres}, but showing the mean and standard errors 
in the residuals for only those Landolt standards observed thrice or more. 
Note the lack of any overall trends with color, although individual stars 
have mean residuals as large as 0.02 mag.}
\label{Avres} 
\end{figure}

In Fig.~\ref{Avres}, the same data are shown, but only for the average residuals of 
objects that have been observed three or more times. The error bars show the 
standard errors. Note that some objects have standard errors that are much smaller 
than their average residual---possibly indicating that there are small but significant 
differences between the Landolt measurements of these stars and ours. A 
possible source of such differences might be the difference in aperture sizes used. 
The mean deviations are less than $\sim 0.02$ mag.  

There is no systematic trend with color (or brightness---not shown), and the 
differences, if real, are random from star to star. This is a possible indication of the 
inherent uncertainties in using these standard stars.  The worst scatter is in the $I$ 
band, but even there, any systematic color dependent trend exceeding $\sim .01$ 
mag is ruled out.  The same data, and also the mean residuals for stars measured 
less than thrice, are presented in Table~\ref{Landoltstars}. This table also identifies 
exactly which Landolt stars were used.

\subsection{Can differences in the \emph{I} bandpass explain the differences
in \emph{I} band photometry versus Stetson's magnitudes?} \label{Banddiff}

One can ask if red giants in NGC~2419, which are all very metal poor 
($[Fe/H] \sim -2$), produce systematically different response through different
$I$ filters, as compared to the Landolt standards at the same $V-I$ color.
This can be studied by synthetic photometry. The standard $I$ passband 
is described numerically by \citet{Bessell90}, in his Table~2. The actual 
filter and CCD response combination used for data in this paper from $MIMO$
is a close match to this standard passband: we will refer to it as the WIYN
$I$ passband. There is a wide range of $I$ filters being vended by commercial 
suppliers, from  filters that are really on the {\it Johnson} rather than  Landolt 
(based on Cousins) system, to ones touted as ``Bessel" $I$ filters that have 
quite large red extensions. We have taken one such example as a `Bad $I$' 
filter.  From simulations we find that the WIYN $I$ filter deviates from the 
standard filter by $\sim 0.009$ mag (fainter) for giants with $V-I \sim 1.5$ mag 
and  $[Fe/H] = -2.0$ (as compared with giants of the same color with solar
metallicity). The  `Bad' filter shows a deviation of $ \sim 0.025 $ mag in the 
same sense with respect to the standard passband. The sense of the 
differences is correct for explaining the discrepancy between our photometry 
and that of Stetson, but the magnitude of the predicted difference is only 
$0.015$, whereas the observed difference is nearly 3 times larger. While 
we have no detailed information on the filter(s) used by Stetson, it is very 
unlikely that it was as bad a mis-match as the example used here for our 
case study. Thus, this too is a very unlikely source of the discrepancy.

\section{Concluding Remarks}

Despite the caveats raised in \S\ref{Intercompare},  this work is a significant 
contribution towards reconciling the large archive of \textit{HST} imaging in 
broadband filters with the \textit{BVRI} photometric system as realized via 
the Landolt standards \citep{Landolt92}. The targets presented here have 
been observed repeatedly with \textit{WFPC2}, and comparing the instrumental 
photometry from images taken at different times will not only add to what we 
know of the temporal variation in CTE of that instrument, but will allow the 
unambiguous calibration of the photometric characteristics over the lifetime 
of the instrument. In other papers, we will present the confrontation of 
\textit{WFPC2} photometry with data obtained at various times and with different
reduction procedures.

This paper, and our understanding, has benefitted greatly from work done on 
these data by Peter Stetson, and from our subsequent discussions with him. We 
are indebted to Marco Sirianni and collaborators for sharing their results of the 
ACS calibration with us before publication. Support for this work was provided by 
NASA through grant HST-AR-09216.01-A from Space Telescope Science 
Institute, which is operated by the Association of Universities for Research in 
Astronomy, Inc., under NASA contract NAS 5-26555. The WIYN Observatory is 
a joint facility of the University of Wisconsin-Madison, Indiana University, Yale 
University, and the National Optical Astronomy Observatory.

\appendix

\section{The Consequence of Spectral Response Mis-match between Measuring 
Setups} 

Here we examine from first principles, the impact and consequences arising 
from the usually encountered situation where the  
measuring system does not have exactly the same wavelength dependence 
as the system that defined the standards (in this case the latter refers to 
Landolt's original setup). We show that the form of the color equation 
that is usually used does not follow the physical demands, except in the 
case when the difference in frequency/wavelength response is small.

Consider a source with an energy distribution $f_{\nu}$. Let the overall 
(telescope, filter and instrument response) response of the original (or standard) 
system be denoted by $\alpha^{S}_{\nu}$, and that of the measuring system being 
used by $\alpha_{\nu}$. The response of the original system is then
\begin{equation}
S^{S} = \int \alpha^{S}_{\nu} f_{\nu} {\rm d}\nu
\end{equation}
and that of the `current' measuring system is 
\begin{equation}
S = \int \alpha_{\nu} f_{\nu} {\rm d}\nu
\end{equation}
We can write 
\begin{equation}
\alpha_{\nu} = C_{\nu} \alpha^{S}_{\nu}
\end{equation}
where $C_{\nu}$ tracks the changes in response of the measuring system with 
respect to the original standard system. If $\nu_{0}$ is the central frequency of 
the passband under consideration, we can expand $C_{\nu}$ as a Taylor series 
about $\nu_{0}$:
\begin{equation}
C_{\nu} = C_{0} + C_{1}(\nu - \nu_{0}) + C_{2}(\nu - \nu_{0})^{2} + \ldots
\end{equation}
And so we can write:
\begin{equation}
S = C_{0} \int \alpha^{S}_{\nu} f_{\nu} {\rm d}\nu + 
    C_{1} \int \alpha^{S}_{\nu} (\nu - \nu_{0}) f_{\nu} {\rm d}\nu + 
    C_{2} \int \alpha^{S}_{\nu} (\nu - \nu_{0})^{2} f_{\nu} {\rm d}\nu + \ldots
\end{equation}
In magnitudes, the left hand side is the instrumental magnitude on the `current' 
measuring setup, and the first term on right hand side (RHS) is a constant 
(zero-point adjustment) plus the true standard magnitude. In the absence of any 
color dependent terms (i.e.\ further terms on the RHS) we would just get (by 
taking the logarithm):
\begin{equation}
m_{observed} = {\rm Offset} + m_{standard} 
\end{equation}
The $n-th$ moment of the spectral energy distribution about the central wavelength 
of the passband is denoted by $ \mu^{n} $, and given by:
\begin{equation}
 \mu^{n} = \frac{\int \alpha^{S}_{\nu} (\nu - \nu_{0})^{n} f_{\nu} {\rm d}\nu}
 {\int \alpha^{S}_{\nu} f_{\nu} {\rm d}\nu}
\end{equation} 
Equation (A5) can thus be re-written as:
\begin{equation}
 S = C_{0} \left( \int \alpha^{S}_{\nu} f_{\nu} {\rm d}\nu \right) ~~  \left( 1 + c_{1} \mu_{1} + 
   c_{2} \mu_{2} + \ldots \right)
\end{equation}
where $ c_{r} = C_{r}/C_{0} $.

Contrast the above equation, which is derived from physical principles, to the {\it ad 
hoc} equation that is used in practice to describe the  response variation, which is:
\begin{equation}
m_{observed} = {\rm Offset} + m_{standard} + B (color) + C (color)^{2} + \ldots
\end{equation} 
The form of equation (A9) can be strictly derived from (A8) only in the case where 
when $c_{1}\mu_{1} < 1$ and $c_{r} \mu_{r}= 0$ for $r \geq 2$.\footnote{The 
McLaurin series expansion of ${\rm ln}(1+x)$ can then be applied to get the log of 
$(1 + c_{1} \mu_{1})$.} One must also assert that \textit{color} in eqn (A9) is 
proportional to $\mu_{1}$ (which behaves like a change in the effective 
wavelength for the passband).  The failure of either of these conditions will make 
eqn (A9) inadequate (to varying degrees, depending on the specifics of the situation).

It should be evident that the {\it ad hoc} form in common use (eqn A9) is thus 
applicable only  if the two measuring  systems are a very near match. If the 
conditions mentioned above are not satisfied, fitting eqn (A9) is tantamount to 
fitting the wrong function. Depending on the severity of the mis-match, this could 
result in systematic color-dependent errors in the photometry. If the entire range of 
colors is evenly sampled, one should notice an increase in the systematic scatter 
when solving for the night constants using the standard stars, as the severity of the 
mis-match increases. This scatter is systemic in nature, and will not be mitigated 
by observing a large number of standards. If the range of colors sampled by the 
standards is patchy, the fitted solution could `run away' in the unsampled regions 
of color. 

In the presence of out of band leaks, the mis-match could be acute,  since one 
expects the color moments (of progressively higher order) to be large as a 
consequence of the large moment `arms'. In addition, in such situations color will 
not remain proportional to the first moment of the energy distribution. In practice, 
however, for stellar work in particular, the progression of the spectral energy 
distribution (SED) with changing physical parameters in the photosphere is very 
constrained. This is manifest in the way that two color diagrams of stars define 
very constrained loci. For this reason, the standard form of the color equations in 
fact work ``better than they should." However trouble should be anticipated for 
non-stellar sources, and in regions of the spectrum where stellar SEDs have large 
dispersion, e.g.\ at the Balmer Jump. 

The above is the mathematical persuasion that filters being used in combination with 
CCDs to reproduce a photometric system established with photo-cathodes be very 
carefully selected, and that they be free of red leaks. We believe that the set used in 
the study presented in this study is as close an approximation as practical. Further, 
the $V$ and $I$ filters used here are a very good match to the $F555W$ and 
$F814W$ filters used with the WFPC2 on \textit{HST}. 

This analysis strongly urges use of bandpasses that be reliably realized by filters that 
are well bounded so that they work with pan-chromatic detectors. This is in contrast to 
\textit{BVRI}, where some of the bands relied on the detector or atmosphere for blocking, 
since such bandpasses are often hard, if not impossible, to reproduce when using 
detectors with very different response, or when they are used in space above the 
atmosphere. Model spectral energy distributions, such as for isochrones and for composite 
colors of galaxies can be synthesized for any passband. However, the rich empirical legacy 
of \textit{BVRI} observations---for instance of Period-Luminosity relations for Cepheids---will 
not be so easily transferred to a new photometric system, and so the problem of designing 
physical filters and calibrating photometry in this `arcane' system will continue to be of importance.

\clearpage

\clearpage 

\clearpage 

\clearpage 

\clearpage

\clearpage

\clearpage

\clearpage

\clearpage

\clearpage

\clearpage

\clearpage

\begin{deluxetable}{lllccc}
\tabletypesize{\scriptsize}
\tablecaption{Summary of Photometric Observations \label{T1}}
\tablewidth{0pt}
\tablehead{
\colhead{UT Date} & \colhead{Telescope} & \colhead{Targets Observed\tablenotemark{a}} &
\colhead{Bandpasses} & \colhead{Landolt Fields\tablenotemark{b}} &
\colhead{Comment}
}
\startdata
2001 Feb 27 & WIYN 3.5m & NGC~2419, Pal~4 & \textit{BVRI} & Rubin~149, SA101-326 & -- \\
2001 Apr 13 & WIYN 3.5m & Pal~14 & \textit{VI} & PG1633+099 & -- \\
2001 Sep 21 & WIYN 3.5m & NGC~2419 & \textit{VI} & SA92-430 & -- \\
2001 Dec 24 & WIYN 3.5m & NGC~2419 & \textit{VI} & SA95-275, SA98-671 & -- \\
    & & & & Rubin~152 & -- \\
2002 Apr 12 & WIYN 3.5m & Pal~3, Pal~14 & \textit{BVRI} & Rubin~149, PG0918+029 & Photometric only in \\
    & & & & PG1633+099, SA110 & 2nd half of night \\
2002 Nov 13 & WIYN 0.9m & NGC~2419, Pal~3, Pal~4 & \textit{BVRI} & SA92-A, Rubin~149 & -- \\
\enddata
\tablenotetext{a}{Targets were often observed a multiple number of times, with 
varied positions on the field of view and/or at different air-masses}
\tablenotetext{b}{Several stars around indicated object and within the 
field of view are Landolt standards. Each field may have multiple observations
at different airmasses}
\end{deluxetable}
%
\begin{deluxetable}{lccc}
\tabletypesize{\normalsize}
\tablecaption{Summary of Photometric Observations \label{T2}}
\tablewidth{0pt}
\tablehead{
\colhead{Target} & 
\colhead{Passband} & 
\colhead{$\epsilon_{sys}$\tablenotemark{a}} &
\colhead{No. of stars \tablenotemark{b}} \\
 & & \colhead{(mags)} &  
}
\startdata
 NGC 2419 & V & 0.0032 & \phn74 \\
          & I & 0.0058 & \phn43 \\
          & R & 0.0043 & \phn87 \\
          & B & 0.0203 & \phn25 \\
 \\
 Pal 14   & V & 0.0050 & 153 \\
          & I & 0.0057 & 158 \\
          & R & 0.0106 & 262 \\
          & B & 0.0046 & 248 \\
 \\
 Pal 4    & V & 0.0123 & 122 \\
          & I & 0.0072 & 114 \\
          & R & 0.0070 & 122 \\
          & B & 0.0061 & \phn91 \\
\enddata
\tablenotetext{a}{See \S \ref{ERRS} for details}
\tablenotetext{b}{This is the number of objects that have measured magnitudes 
in this passband for {\it all} exposures of this target: see \S \ref{ERRS} }
\end{deluxetable}
\clearpage
%
\begin{deluxetable}{rcrccrrrrrrrrrrrr}
\tabletypesize{\tiny}
\rotate
\tablecaption{Stars in the NGC2419 Field with best Photometry  \label{N2419best}}
\tablewidth{0pt}
\tablehead{
\colhead{Star ID} &
\colhead{X} & \colhead{Y} & \colhead{R.A.} & \colhead{Decl.} & 
\colhead{$<V>$} & \colhead{$\epsilon(V)$} & \colhead{$n_V$}  &
\colhead{$<I>$} & \colhead{$\epsilon(I)$} & \colhead{$n_I$}  &
\colhead{$<R>$} & \colhead{$\epsilon(R)$} & \colhead{$n_R$}  &
\colhead{$<B>$} & \colhead{$\epsilon(B)$} & \colhead{$n_B$}  \\
 & \colhead{(pix)} & \colhead{(pix)} & \colhead{(J2000)} & \colhead{(J2000)} &
}
\startdata
       1 &      266.104 &  3962.359 &  7:37:53.69 & 38:59:07.0 &  12.808 &  0.008 &   4  &   12.146 &  0.012 &   4  &  12.447 &  0.005 &   4  &  13.291 &  0.013 &  3 \\
       2 &      447.962 &  3334.508 &  7:38:01.27 & 38:58:41.6 &  16.065 &  0.007 &   5  &   15.173 &  0.007 &   5  &  15.608 &  0.006 &   4  &  16.810 &  0.010 &  4 \\
       3 &      764.747 &  1553.491 &  7:38:22.74 & 38:57:57.4 &  17.518 &  0.005 &   6  &   16.972 &  0.012 &   5  &  17.258 &  0.010 &   3  &  17.860 &  0.010 &  5 \\
       4 &      960.799 &  2767.838 &  7:38:08.13 & 38:57:29.7 &  17.209 &  0.007 &   7  &   16.169 &  0.006 &   6  &  16.671 &  0.008 &   5  &  18.113 &  0.013 &  8 \\
       5 &     1068.558 &  2806.926 &  7:38:07.67 & 38:57:14.5 &  17.792 &  0.007 &   8  &   17.126 &  0.008 &   7  &  17.455 &  0.011 &   6  &  18.300 &  0.007 &  7 \\
       6 &     1094.433 &  2071.915 &  7:38:16.52 & 38:57:11.0 &  18.012 &  0.008 &   8  &   15.909 &  0.012 &   7  &  17.024 &  0.012 &   6  &  19.581 &  0.004 &  8 \\
       7 &     1103.414 &  2972.147 &  7:38:05.69 & 38:57:09.6 &  16.857 &  0.009 &   7  &   15.833 &  0.006 &   6  &  16.316 &  0.009 &   5  &  17.724 &  0.005 &  8 \\
       8 &     1204.976 &  2728.613 &  7:38:08.63 & 38:56:55.4 &  16.084 &  0.014 &   5  &   15.341 &  0.006 &   5  &  15.708 &  0.009 &   6  &  16.685 &  0.011 &  6 \\
\enddata
\end{deluxetable}
\clearpage
%
\begin{deluxetable}{rrccrrrrrrrrrrrrrrrr}
\tabcolsep=5pt
\tabletypesize{\tiny}
\rotate
\tablecaption{Stars in the NGC2419 Field with at least $V$ and $I$ Photometry\tablenotemark{a}\label{N2419VIexist}}
\tablewidth{0pt}
\tablehead{
\colhead{X} & \colhead{Y} & \colhead{R.A.} & \colhead{Decl.} & 
\colhead{$<V>$} & \colhead{$\epsilon(V)$} & \colhead{$n_V$} & \colhead{$\chi^{2}_{\nu}(V)$} &
\colhead{$<I>$} & \colhead{$\epsilon(I)$} & \colhead{$n_I$} & \colhead{$\chi^{2}_{\nu}(I)$} &
\colhead{$<R>$} & \colhead{$\epsilon(R)$} & \colhead{$n_R$} & \colhead{$\chi^{2}_{\nu}(R)$} &
\colhead{$<B>$} & \colhead{$\epsilon(B)$} & \colhead{$n_B$} & \colhead{$\chi^{2}_{\nu}(B)$} \\
\colhead{(pix)} & \colhead{(pix)} & \colhead{(J2000)} & \colhead{(J2000)} &
 & & & &  & & & &  & & & &  & & & 
}
\startdata
      12.084 &  1063.137 &  7:38:28.60 & 38:59:43.2 &  16.207 &  0.021 &  2  &   21.37  &   15.446 &  0.004 &  2  &    0.13 &   ---   &   ---  &  0 &    ---  &  16.812 & 0.034 &  2 &   49.13 \\
      42.603 &  2613.932 &  7:38:09.91 & 38:59:38.8 &  13.195 &  0.010 &  4  &    9.24  &   12.518 &  0.005 &  4  &    1.14 &  12.829 &  0.004 &  2 &    0.59 &  13.625 & 0.018 &  4 &   34.14 \\
      53.641 &  2073.155 &  7:38:16.43 & 38:59:37.3 &  19.744 &  0.010 &  4  &    1.09  &   18.686 &  0.019 &  4  &    0.27 &   ---   &   ---  &  0 &    ---  &  20.553 & 0.048 &  2 &    2.64 \\
      70.817 &  1355.050 &  7:38:25.08 & 38:59:34.9 &  14.686 &  0.014 &  2  &    9.97  &   14.041 &  0.006 &  3  &    0.51 &   ---   &   ---  &  0 &    ---  &  15.172 & 0.017 &  3 &    8.74 \\
      88.735 &  3234.700 &  7:38:02.44 & 38:59:32.2 &  18.979 &  0.013 &  4  &    2.06  &   18.279 &  0.036 &  4  &    1.57 &  18.582 &  0.032 &  3 &    1.26 &  19.410 & 0.146 &  2 &    6.96 \\
      94.668 &  1579.926 &  7:38:22.37 & 38:59:31.6 &  19.797 &  0.010 &  4  &    1.16  &   18.776 &  0.018 &  4  &    0.47 &   ---   &   ---  &  0 &    ---  &  20.627 & 0.080 &  2 &    7.63 \\
     100.635 &   271.525 &  7:38:38.14 & 38:59:30.8 &  16.703 &  0.031 &  2  &   47.05  &   15.812 &  0.012 &  2  &    6.76 &   ---   &   ---  &  0 &    ---  &   ---   &  ---  &  1 &    ---  \\
     139.994 &   489.142 &  7:38:35.52 & 38:59:25.2 &  17.472 &  0.026 &  2  &   28.02  &   16.954 &  0.009 &  2  &    0.55 &   ---   &   ---  &  0 &    ---  &  17.845 & 0.054 &  2 &   32.40 \\
\enddata
\tablenotetext{a}{The full text of this table may be found at: ftp://taurus.tuc.noao.edu/pub/saha/Photseq/table4.txt}
\end{deluxetable}
\clearpage
\begin{deluxetable}{lccccrrrrrrrrrrrrrrrr}
\tabletypesize{\tiny}
\tabcolsep=4pt
\rotate
\tablecaption{Likely Variable Stars in the NGC2419 Field \label{N2419variab}}
\tablewidth{0pt}
\tablehead{
\colhead{Star ID} &
\colhead{X} & \colhead{Y} & \colhead{R.A.} & \colhead{Decl.} & 
\colhead{$<V>$} & \colhead{$\epsilon(V)$} & \colhead{$n_V$} & \colhead{$\chi^{2}_{\nu}(V)$} &
\colhead{$<I>$} & \colhead{$\epsilon(I)$} & \colhead{$n_I$} & \colhead{$\chi^{2}_{\nu}(I)$} &
\colhead{$<R>$} & \colhead{$\epsilon(R)$} & \colhead{$n_R$} & \colhead{$\chi^{2}_{\nu}(R)$} &
\colhead{$<B>$} & \colhead{$\epsilon(B)$} & \colhead{$n_B$} & \colhead{$\chi^{2}_{\nu}(B)$} \\
 & \colhead{(pix)} & \colhead{(pix)} & \colhead{(J2000)} & \colhead{(J2000)} &
 & & & &  & & & &  & & & &  & & & 
}

\startdata
   V1  &    1495.028 &  2459.969 &  7:38:11.88 & 38:56:14.6 &  20.410 &  0.030 &  5  &   20.50  &   20.037 &  0.063 &  6  &   52.35 &  20.217 & 0.036 &  5 &   30.42 &  20.665 & 0.028 &  5 &   18.48 \\
   V2  &    1829.114 &  2584.987 &  7:38:10.40 & 38:55:27.7 &  20.408 &  0.082 &  6  &  173.13  &   20.028 &  0.071 &  4  &   65.38 &  20.298 & 0.073 &  6 &   84.29 &  20.949 & 0.063 &  7 &   88.39 \\
   V3  &    1858.845 &  2224.339 &  7:38:14.75 & 38:55:23.5 &  21.204 &  0.023 &  7  &    5.23  &   20.381 &  0.092 &  7  &   41.93 &  20.831 & 0.070 &  6 &   36.61 &  21.722 & 0.031 &  3 &    1.32 \\
   V4  &    2171.801 &  1975.868 &  7:38:17.76 & 38:54:39.6 &  20.474 &  0.073 & 10  &  244.39  &   19.831 &  0.047 &  9  &   59.76 &  19.989 & 0.077 & 11 &  396.75 &  20.699 & 0.089 & 13 &  359.09 \\
   V5  &    2172.556 &  1806.291 &  7:38:19.80 & 38:54:39.5 &  19.922 &  0.123 & 11  & 1180.60  &   19.481 &  0.075 & 10  &  194.65 &  20.189 & 0.035 & 12 &   38.10 &  20.916 & 0.057 & 12 &  139.70 \\
   V6  &    2309.284 &  2978.859 &  7:38:05.70 & 38:54:20.1 &  20.485 &  0.064 & 11  &  168.63  &   19.895 &  0.060 & 10  &   41.90 &  20.269 & 0.041 &  7 &   27.07 &  21.039 & 0.040 &  8 &   50.80 \\
   V7  &    2328.558 &  2106.259 &  7:38:16.21 & 38:54:17.5 &  20.653 &  0.082 &  8  &  130.97  &   20.042 &  0.061 &  8  &   53.15 &  20.299 & 0.051 & 11 &   26.83 &  21.192 & 0.100 & 11 &   71.34 \\
   V8  &    2407.928 &  2585.301 &  7:38:10.45 & 38:54:06.3 &  20.767 &  0.026 &  6  &    3.71  &   19.864 &  0.053 &  4  &   34.73 &  20.357 & 0.058 &  7 &   62.93 &  21.389 & 0.020 & 10 &    2.71 \\
\enddata
\end{deluxetable}
\clearpage
%
\begin{deluxetable}{rccccrrcrrcrrcrrc}
\tabletypesize{\tiny}
\rotate
\tablecaption{Stars in the Pal 4 Field with best Photometry  \label{Pal4best}}
\tablewidth{0pt}
\tablehead{
\colhead{Star ID} &
\colhead{X} & \colhead{Y} & \colhead{R.A.} & \colhead{Decl.} & 
\colhead{$<V>$} & \colhead{$\epsilon(V)$} & \colhead{$n_V$}  &
\colhead{$<I>$} & \colhead{$\epsilon(I)$} & \colhead{$n_I$}  &
\colhead{$<R>$} & \colhead{$\epsilon(R)$} & \colhead{$n_R$}  &
\colhead{$<B>$} & \colhead{$\epsilon(B)$} & \colhead{$n_B$}  \\
 & \colhead{(pix)} & \colhead{(pix)} & \colhead{(J2000)} & \colhead{(J2000)} &
 & & &  & & &   & & &   & &  
}
\startdata
     1 &   1392.870 &  2104.530 &  11:29:17.58 & 28:55:22.7 &  21.322 &  0.014 &  5  &   20.378 &  0.013 &   5  &  20.859 &  0.010 &   5  &  22.090 &  0.009 &  5 \\
     2 &   1586.340 &  1492.180 &  11:29:11.00 & 28:55:49.7 &  20.930 &  0.008 &  6  &   18.218 &  0.013 &   6  &  19.728 &  0.014 &   5  &  22.421 &  0.010 &  5 \\
     3 &   1917.170 &   598.700 &  11:29:01.41 & 28:56:35.9 &  18.056 &  0.011 &  6  &   17.898 &  0.009 &   6  &  17.984 &  0.011 &   6  &  18.170 &  0.013 &  6 \\
     4 &   2124.650 &  2354.410 &  11:29:20.23 & 28:57:05.6 &  20.737 &  0.011 &  6  &   19.946 &  0.009 &   6  &  20.335 &  0.012 &   6  &  21.309 &  0.013 &  6 \\
     5 &   2246.010 &  2135.150 &  11:29:17.87 & 28:57:22.6 &  21.124 &  0.012 &  6  &   20.165 &  0.014 &   6  &  20.654 &  0.014 &   6  &  21.900 &  0.014 &  6 \\
     6 &   2306.050 &  1567.040 &  11:29:11.78 & 28:57:30.9 &  20.164 &  0.008 &  6  &   19.156 &  0.005 &   6  &  19.657 &  0.007 &   6  &  21.065 &  0.012 &  6 \\
     7 &   2406.070 &  2079.100 &  11:29:17.27 & 28:57:45.1 &  19.800 &  0.004 &  6  &   18.740 &  0.007 &   6  &  19.256 &  0.003 &   6  &  20.743 &  0.004 &  6 \\
     8 &   2410.330 &  2138.220 &  11:29:17.90 & 28:57:45.7 &  20.698 &  0.014 &  6  &   19.925 &  0.014 &   6  &  20.305 &  0.011 &   6  &  21.291 &  0.009 &  6 \\
\enddata
\end{deluxetable}
\clearpage
%
\begin{deluxetable}{rrccrrrrrrrrrrrrrrrr}
\tabletypesize{\tiny}
\tabcolsep=5pt
\rotate
\tablecaption{Stars in the Pal 4 Field with at least $V$ and $I$ Photometry\tablenotemark{a}\label{Pal4VIexist}}
\tablewidth{0pt}
\tablehead{
\colhead{X} & \colhead{Y} & \colhead{R.A.} & \colhead{Decl.} & 
\colhead{$<V>$} & \colhead{$\epsilon(V)$} & \colhead{$n_V$} & \colhead{$\chi^{2}_{\nu}(V)$} &
\colhead{$<I>$} & \colhead{$\epsilon(I)$} & \colhead{$n_I$} & \colhead{$\chi^{2}_{\nu}(I)$} &
\colhead{$<R>$} & \colhead{$\epsilon(R)$} & \colhead{$n_R$} & \colhead{$\chi^{2}_{\nu}(R)$} &
\colhead{$<B>$} & \colhead{$\epsilon(B)$} & \colhead{$n_B$} & \colhead{$\chi^{2}_{\nu}(B)$} \\
\colhead{(pix)} & \colhead{(pix)} & \colhead{(J2000)} & \colhead{(J2000)} &
 & & & &  & & & &  & & & &  & & & 
}
\startdata
     108.190 &  1539.510 &  11:29:11.56 & 28:52:21.9 &  23.591 &  0.127 &  2 &    2.79  &   21.883 &  0.091 &  2  &    2.74 &   ---   &   ---  &  1 &    ---   &   ---   &   ---  & 0 &   ---   \\
     130.000 &  3599.400 &  11:29:33.64 & 28:52:25.4 &  21.775 &  0.074 &  3 &   18.02  &   21.510 &  0.341 &  2  &   32.58 &  21.393 &  0.056 &  3 &    10.16 &   ---   &   ---  & 1 &   ---   \\
     235.250 &  3257.850 &  11:29:29.97 & 28:52:40.2 &  22.101 &  0.041 &  3 &    5.11  &   20.560 &  0.035 &  3  &    3.64 &  21.251 &  0.046 &  3 &    22.64 &   ---   &   ---  & 0 &   ---   \\
     363.150 &  3061.880 &  11:29:27.87 & 28:52:58.1 &  22.006 &  0.084 &  3 &   21.22  &   21.360 &  0.042 &  3  &    1.36 &  21.681 &  0.080 &  3 &    18.09 &   ---   &   ---  & 1 &   ---   \\
     420.510 &  3754.630 &  11:29:35.29 & 28:53:06.3 &  23.367 &  0.069 &  2 &    0.11  &   22.502 &  0.131 &  2  &    1.75 &  23.017 &  0.055 &  2 &     0.62 &   ---   &   ---  & 0 &   ---   \\
     480.930 &  1991.220 &  11:29:16.39 & 28:53:14.4 &  22.905 &  0.044 &  2 &    0.24  &   20.055 &  0.031 &  2  &   10.87 &  21.669 &  0.019 &  2 &     0.21 &   ---   &   ---  & 0 &   ---   \\
     487.270 &  2333.120 &  11:29:20.06 & 28:53:15.4 &  20.262 &  0.019 &  3 &    8.18  &   17.045 &  0.023 &  3  &   65.23 &  18.796 &  0.015 &  4 &    21.62 &  21.911 &  0.068 & 3 &   13.34 \\
     502.440 &  1326.380 &  11:29:09.27 & 28:53:17.3 &  23.835 &  0.104 &  2 &    0.61  &   22.388 &  0.083 &  2  &    0.85 &   ---   &   ---  &  1 &    ---   &   ---   &   ---  & 0 &   ---   \\
\enddata
\tablenotetext{a}{The full text of this table may be found at: ftp://taurus.tuc.noao.edu/pub/saha/Photseq/table7.txt}
\end{deluxetable}
\clearpage
%
\begin{deluxetable}{cccccrrrrrrrrrrrrrrrr}
\tabletypesize{\tiny}
\tabcolsep=4pt
\rotate
\tablecaption{Likely Variable Stars in the Pal~4 Field \label{Pal4variab}}
\tablewidth{0pt}
\tablehead{
\colhead{Star ID} &
\colhead{X} & \colhead{Y} & \colhead{R.A.} & \colhead{Decl.} & 
\colhead{$<V>$} & \colhead{$\epsilon(V)$} & \colhead{$n_V$} & \colhead{$\chi^{2}_{\nu}(V)$} &
\colhead{$<I>$} & \colhead{$\epsilon(I)$} & \colhead{$n_I$} & \colhead{$\chi^{2}_{\nu}(I)$} &
\colhead{$<R>$} & \colhead{$\epsilon(R)$} & \colhead{$n_R$} & \colhead{$\chi^{2}_{\nu}(R)$} &
\colhead{$<B>$} & \colhead{$\epsilon(B)$} & \colhead{$n_B$} & \colhead{$\chi^{2}_{\nu}(B)$} \\
  & \colhead{(pix)} & \colhead{(pix)} & \colhead{(J2000)} & \colhead{(J2000)} &
 & & & &  & & & &  & & & &  & & & 
}
\startdata
  V1  &    2014.840 &  2730.520 & 11:29:24.27 & 28:56:50.3 &  22.185 &  0.190 &  5  &   94.03  &   21.503 &  0.142 &  6  &   25.96 &  21.824 & 0.168 &  5 &  135.51 &  22.697 & 0.200 &  5 &  152.63 \\
  V2  &    2609.960 &  1846.370 & 11:29:14.77 & 28:58:13.7 &  17.858 &  0.436 &  4  & 30355.61  &   15.580 &  0.204 &  4  & 6679.30 &  16.611 & 0.403 &  4 & 26047.48 &  19.840 & 0.383 &  6 & 17645.34 \\
  V3  &    2699.190 &  2092.920 & 11:29:17.41 & 28:58:26.3 &  20.258 &  0.075 &  5  &  228.70  &   19.657 &  0.061 &  6  &   98.17 &  19.823 & 0.076 &  5 &  374.85 &  20.641 & 0.096 &  6 &  582.93 \\
\enddata
\end{deluxetable}
\clearpage
%
\begin{deluxetable}{rrrccrrcrrcrrcrrc}
\tabletypesize{\tiny}
\rotate
\tablecaption{Stars in the Pal 14 Field with best Photometry  \label{Pal14best}}
\tablewidth{0pt}
\tablehead{
\colhead{Star ID} &
\colhead{X} & \colhead{Y} & \colhead{R.A.} & \colhead{Decl.} & 
\colhead{$<V>$} & \colhead{$\epsilon(V)$} & \colhead{$n_V$}  &
\colhead{$<I>$} & \colhead{$\epsilon(I)$} & \colhead{$n_I$}  &
\colhead{$<R>$} & \colhead{$\epsilon(R)$} & \colhead{$n_R$}  &
\colhead{$<B>$} & \colhead{$\epsilon(B)$} & \colhead{$n_B$}  \\
 & \colhead{(pix)} & \colhead{(pix)} & \colhead{(J2000)} & \colhead{(J2000)} &
 & & &  & & &   & & &   & &  
}
\startdata
     1 &      30.520 &  3272.550 &  16:11:11.10 & 14:53:54.2 &  19.754 &  0.008 &  3  &   18.389 &  0.004 &   3  &  19.012 &  0.015 &   4  &  20.821 &  0.009 &  4 \\
     2 &      52.970 &  3075.740 &  16:11:09.19 & 14:53:57.3 &  18.655 &  0.005 &  3  &   17.933 &  0.004 &   4  &  18.275 &  0.006 &   6  &  19.142 &  0.006 &  4 \\
     3 &      58.800 &  2045.850 &  16:10:59.18 & 14:53:57.7 &  20.697 &  0.008 &  4  &   19.728 &  0.010 &   4  &  20.203 &  0.009 &   6  &  21.462 &  0.012 &  5 \\
     4 &      65.000 &   653.680 &  16:10:45.66 & 14:53:58.1 &  18.282 &  0.003 &  3  &   17.332 &  0.005 &   3  &  17.773 &  0.006 &   4  &  19.144 &  0.006 &  3 \\
     5 &     257.770 &  2179.340 &  16:11:00.48 & 14:54:25.8 &  19.669 &  0.003 &  4  &   18.897 &  0.006 &   4  &  19.269 &  0.003 &   6  &  20.315 &  0.005 &  5 \\
     6 &     327.040 &  3781.770 &  16:11:16.04 & 14:54:36.1 &  20.099 &  0.006 &  3  &   19.292 &  0.007 &   3  &  19.663 &  0.014 &   4  &  20.666 &  0.006 &  3 \\
     7 &     372.780 &  2580.600 &  16:11:04.37 & 14:54:42.1 &  18.112 &  0.007 &  4  &   17.260 &  0.005 &   4  &  17.667 &  0.005 &   6  &  18.793 &  0.003 &  5 \\
     8 &     466.250 &  2052.810 &  16:10:59.24 & 14:54:55.0 &  19.091 &  0.005 &  4  &   18.337 &  0.007 &   4  &  18.711 &  0.004 &   6  &  19.676 &  0.004 &  5 \\
\enddata
\end{deluxetable}
\clearpage
%
\begin{deluxetable}{rrccrrrrrrrrrrrrrrrr}
\tabletypesize{\tiny}
\tabcolsep=5pt
\rotate
\tablecaption{Stars in the Pal 14 Field with at least $V$ and $I$ Photometry\tablenotemark{a}\label{Pal14VIexist}}
\tablewidth{0pt}
\tablehead{
\colhead{X} & \colhead{Y} & \colhead{R.A.} & \colhead{Decl.} & 
\colhead{$<V>$} & \colhead{$\epsilon(V)$} & \colhead{$n_V$} & \colhead{$\chi^{2}_{\nu}(V)$} &
\colhead{$<I>$} & \colhead{$\epsilon(I)$} & \colhead{$n_I$} & \colhead{$\chi^{2}_{\nu}(I)$} &
\colhead{$<R>$} & \colhead{$\epsilon(R)$} & \colhead{$n_R$} & \colhead{$\chi^{2}_{\nu}(R)$} &
\colhead{$<B>$} & \colhead{$\epsilon(B)$} & \colhead{$n_B$} & \colhead{$\chi^{2}_{\nu}(B)$} \\
\colhead{(pix)} & \colhead{(pix)} & \colhead{(J2000)} & \colhead{(J2000)} &
 & & & &  & & & &  & & & &  & & & 
}
\startdata
      12.830 &  3762.790 & 16:11:15.86 & 14:53:51.9 &  21.336 &  0.013 &  3  &    0.98  &   20.342 &  0.015 &  3  &    1.72 &  20.801 &  0.029 &  2 &    9.51 &  21.980 & 0.018 &  2 &    0.25 \\
      30.520 &  3272.550 & 16:11:11.10 & 14:53:54.2 &  19.754 &  0.008 &  3  &    4.27  &   18.389 &  0.004 &  3  &    0.80 &  19.012 &  0.015 &  4 &   15.50 &  20.821 & 0.009 &  4 &    1.86 \\
      52.790 &  1909.790 & 16:10:57.86 & 14:53:56.9 &  23.437 &  0.062 &  2  &    1.28  &   20.886 &  0.036 &  3  &    6.20 &  22.407 &  0.030 &  5 &    3.42 &   ---   &  ---  &  1 &    ---  \\
      52.970 &  3075.740 & 16:11:09.19 & 14:53:57.3 &  18.655 &  0.005 &  3  &    2.36  &   17.933 &  0.004 &  4  &    2.16 &  18.275 &  0.006 &  6 &    6.91 &  19.142 & 0.006 &  4 &    4.04 \\
      54.100 &  3171.720 & 16:11:10.12 & 14:53:57.5 &  21.087 &  0.010 &  3  &    2.45  &   18.451 &  0.025 &  4  &   81.22 &  19.867 &  0.010 &  6 &   10.47 &  22.800 & 0.024 &  4 &    1.90 \\
      57.040 &  1371.220 & 16:10:52.63 & 14:53:57.2 &  23.688 &  0.076 &  2  &    0.03  &   22.686 &  0.169 &  3  &    4.09 &  23.097 &  0.116 &  2 &    4.81 &   ---   &  ---  &  1 &    ---  \\
      57.190 &  2776.150 & 16:11:06.28 & 14:53:57.8 &  23.279 &  0.053 &  3  &    1.22  &   22.348 &  0.081 &  3  &    1.66 &  23.025 &  0.065 &  5 &    4.89 &  23.819 & 0.072 &  4 &    2.53 \\
      58.800 &  2045.850 & 16:10:59.18 & 14:53:57.7 &  20.697 &  0.008 &  4  &    1.90  &   19.728 &  0.010 &  4  &    3.13 &  20.203 &  0.009 &  6 &    3.29 &  21.462 & 0.012 &  5 &    3.50 \\
\enddata
\tablenotetext{a}{The full text of this table may be found at: ftp://taurus.tuc.noao.edu/pub/saha/Photseq/table10.txt}
\end{deluxetable}
\clearpage
%
\begin{deluxetable}{crrccrrrrrrrrrrrrrrrr}
\tabletypesize{\tiny}
\tabcolsep=4pt
\rotate
\tablecaption{Likely Variable Stars in the Pal~14 Field\label{Pal14variab}}
\tablewidth{0pt}
\tablehead{
\colhead{Star ID} &
\colhead{X} & \colhead{Y} & \colhead{R.A.} & \colhead{Decl.} & 
\colhead{$<V>$} & \colhead{$\epsilon(V)$} & \colhead{$n_V$} & \colhead{$\chi^{2}_{\nu}(V)$} &
\colhead{$<I>$} & \colhead{$\epsilon(I)$} & \colhead{$n_I$} & \colhead{$\chi^{2}_{\nu}(I)$} &
\colhead{$<R>$} & \colhead{$\epsilon(R)$} & \colhead{$n_R$} & \colhead{$\chi^{2}_{\nu}(R)$} &
\colhead{$<B>$} & \colhead{$\epsilon(B)$} & \colhead{$n_B$} & \colhead{$\chi^{2}_{\nu}(B)$} \\
  & \colhead{(pix)} & \colhead{(pix)} & \colhead{(J2000)} & \colhead{(J2000)} &
 & & & &  & & & &  & & & &  & & & 
}
\startdata
  V1   &     750.640 &  3667.320 & 16:11:14.92 & 14:55:35.6 &  23.382 &  0.051 &  6  &     1.71  &   21.336 &  0.078 &  6  &   23.62 &  22.389 & 0.043 &  4 &     7.03 & --10.000 & --1.000 & 1 &  --100.00 \\
  V2   &     776.070 &  1393.620 & 16:10:52.83 & 14:55:38.3 &  19.869 &  0.233 &  6  &  4969.24  &   18.881 &  0.034 &  6  &   95.32 &  19.357 & 0.005 &  4 &     2.88 &  20.700 &  0.013 & 3 &     4.93 \\
  V3   &     841.310 &  1739.040 & 16:10:56.18 & 14:55:47.6 &  19.866 &  0.068 &  6  &   436.13  &   19.241 &  0.010 &  6  &    5.52 &  19.554 & 0.005 &  6 &     3.03 &  20.341 &  0.015 & 5 &    10.00 \\
  V4   &     888.810 &   883.730 & 16:10:47.87 & 14:55:54.0 &  20.943 &  0.100 &  6  &   240.46  &   20.220 &  0.034 &  6  &   32.05 &  20.560 & 0.005 &  6 &     0.45 &  21.587 &  0.007 & 5 &     1.17 \\
  V5   &    1334.060 &  1962.690 & 16:10:58.34 & 14:56:57.0 &  21.155 &  0.051 &  9  &    94.90  &   19.843 &  0.005 &  9  &    1.69 &  20.441 & 0.014 &  6 &     6.72 &  22.200 &  0.031 & 5 &     6.69 \\
  V6   &    1487.820 &  1992.960 & 16:10:58.63 & 14:57:18.6 &  21.568 &  0.062 &  6  &    34.19  &   21.307 &  0.024 &  4  &    1.97 &  21.501 & 0.073 &  6 &    89.40 &  21.849 &  0.047 & 4 &    29.80 \\
  V7   &    1867.460 &  2161.920 & 16:11:00.26 & 14:58:12.0 &  21.315 &  0.022 &  8  &    13.13  &   20.546 &  0.062 &  8  &   44.30 &  20.881 & 0.020 &  4 &     8.61 &  22.033 &  0.045 & 4 &     8.70 \\
\enddata
\end{deluxetable}
\clearpage
%
\begin{deluxetable}{lrrrr}
\tabletypesize{\small}
\tablecaption{Mean Offsets of Various Photometry of NGC~2419 vs.\ this paper \label{Comptable}}
\tablewidth{0pt}
\tablehead{
\colhead{Study} & \multicolumn{4}{c}{Mean Offsets (Study - this paper)\tablenotemark{a}} \\
 & \colhead{$~~~B~~~$} & \colhead{$~~~V~~~$} & \colhead{$~~~R~~~$} & \colhead{$~~~I~~~$}
}
\startdata
Stetson sequence    & +.014   & $-$.014   & $-$.017   & $-$.038  \\
                    & (.021\rlap{)}  & (.004\rlap{)}  & (.005\rlap{)}  & (.006\rlap{)} \\
 \\
Davis sequence      & +.027   &$-$.005   & +.001   & $-$.022  \\
                    & (.021\rlap{)}  & (.005\rlap{)}  & (.007\rlap{)}  & (.007\rlap{)} \\
 \\
Dolphin WFPC2\tablenotemark{b}  &   ---   & +.020   &   ---   & $-$.011  \\
                    &         & (.006\rlap{)}  &         & (.007\rlap{)} \\
 \\
Dolphin's independent &   ---   & +.002   &   ---   & +.003  \\ 
calib of WIYN data   &         & (.004\rlap{)}  &         & (.007\rlap{)} \\
 \\
WIYN 0.9m data only & +.055   & $-$.002   & +.021   & +.032  \\
                    & (.031\rlap{)}  & (.012\rlap{)}  & (.009\rlap{)}  & (.018\rlap{)} \\ 
\\
Sirianni et al.\ $ACS$\tablenotemark{c} & \multicolumn{4}{c}{Offsets within $\pm  0.02$ in $V$, $R$, and $I$} \\
\enddata
\tablenotetext{a} {Values in parentheses indicate errors in the mean 
differences of the stars compared, to which the systematic error estimates 
from Table~\ref{T2} have been added in quadrature. These may be slight 
underestimates, since systematic error estimates for the sequences being 
compared to, which should also 
be added in quadrature, are not generally available.}
\tablenotetext{b} {The calibration is based on WFPC2 observations of $\omega$ Cen, and the photometric sequence by \citet{Walker94}.}
\tablenotetext{c} {Estimates are from privately communicated figures. Numbers not available}
\end{deluxetable}
\clearpage
%
\begin{deluxetable}{lrrrrrrrrrrrrrrrr}
\tabletypesize{\tiny}
\rotate
\tablecaption{Mean Residuals and Standard Errors of the Calibrating Landolt Standard stars \label{Landoltstars}} 
\tablewidth{0pt}
\tablehead{
\colhead{Object}&
\colhead{V} & \colhead{$B-V$} & \colhead{$V-R$} & \colhead{$R-I$} &
\colhead{$\Delta V$\tablenotemark{a}} & \colhead{$\epsilon_{V}$} & \colhead{$n(V)$} &
\colhead{$\Delta I$\tablenotemark{a}} & \colhead{$\epsilon_{I}$} & \colhead{$n(I)$} &
\colhead{$\Delta B$\tablenotemark{a}} & \colhead{$\epsilon_{B}$} & \colhead{$n(B)$} &
\colhead{$\Delta R$\tablenotemark{a}} & \colhead{$\epsilon_{R}$} & \colhead{$n(R)$} 
}
\startdata
     SA101-262  &  14.295  &  0.784  &  0.440  &  0.387  &   0.008 &   ---   &   2     &   0.005 &   ---   &   2    &    0.001 &   ---    &  2    &    0.007 &   ---    &   2  \\       
     SA101-268  &  14.380  &  1.531  &  1.040  &  1.200  &   0.005 &   ---   &   2     &   0.021 &   ---   &   2    &    0.017 &   ---    &  2    &    0.009 &   ---    &   2  \\       
     SA101-326  &  14.923  &  0.729  &  0.406  &  0.375  &  $-$0.010 &   ---   &   2     &  $-$0.022 &   ---   &   2    &   $-$0.013 &   ---    &  2    &   $-$0.010 &   ---    &   2  \\      
     SA101-330  &  13.723  &  0.577  &  0.346  &  0.338  &   0.004 &   ---   &   2     &   0.001 &   ---   &   2    &   $-$0.002 &   ---    &  2    &    0.007 &   ---    &   2  \\      
     SA104-330  &  15.296  &  0.594  &  0.369  &  0.371  &   0.001 &   0.006 &  11     &   0.012 &   0.003 &  11    &   $-$0.024 &   0.003  &  8    &    0.007 &   0.004  &   8  \\      
     SA104-333  &  15.459  &  0.832  &  0.476  &  0.374  &  $-$0.020 &   ---   &   1     &   0.102 &   ---   &   1    &    ---   &   ---    &  0    &    ---   &   ---    &   0  \\      
     SA104-334  &  13.484  &  0.518  &  0.323  &  0.331  &   0.004 &   0.002 &   9     &  $-$0.008 &   0.002 &   9    &    0.003 &   0.002  &  9    &    0.001 &   0.002  &   9  \\      
     SA104-336  &  14.404  &  0.830  &  0.461  &  0.403  &   0.002 &   0.003 &  10     &   0.004 &   0.003 &  10    &    0.009 &   0.002  &  6    &   $-$0.001 &   0.003  &   6  \\
\enddata
\tablenotetext{a}{The sense of $\Delta mag$ is ``Landolt value" minus ``Mean from WIYN 3.5m observations"}
\end{deluxetable}
\end{document}